\documentclass[AMA,LATO1COL]{WileyNJD-v2}
\usepackage{moreverb}
\usepackage{longtable}
\usepackage{multirow}
\usepackage{multicol}
\usepackage[flushleft]{threeparttable}
\usepackage{xspace}
\usepackage{pifont}
\usepackage{enumitem}
\usepackage{subfigure}
\usepackage{pbox}
\usepackage[skins,xparse,breakable]{tcolorbox}
\usepackage{CJK}
\usepackage{amsmath}
\usepackage{amssymb}

\makeatletter
\DeclareRobustCommand\onedot{\futurelet\@let@token\@onedot}
\def\@onedot{\ifx\@let@token.\else.\null\fi\xspace}
\def\eg{\emph{e.g}\onedot} \def\Eg{\emph{E.g}\onedot}
\def\ie{\emph{i.e}\onedot} 
 
\def\etc{\emph{etc}\onedot} 
 
\def\etal{\emph{et al}\onedot}

\makeatother

\newcommand\BibTeX{{\rmfamily B\kern-.05em \textsc{i\kern-.025em b}\kern-.08em
T\kern-.1667em\lower.7ex\hbox{E}\kern-.125emX}}

\def\modelname{WELL\xspace}
\def\modelnameE{WELL-1\xspace}
\def\modelnameF{\textit{\textbf{WE}}akly supervised bug \textit{\textbf{L}}oca\textit{\textbf{L}}ization\xspace}

\def\Dvar{VarMisuse\xspace}
\def\Dbiop{BiOpMisuse\xspace}
\def\Dbound{BoundError\xspace}
\def\Dstu{StuBug\xspace}
\makeatother

\newcommand{\modelnameL}[1]{WELL-{#1}\xspace}

\renewcommand{\paragraph}[1]{\vskip 0.05in \noindent {\bf #1.}}

\articletype{Article Type}%

\received{<day> <Month>, <year>}
\revised{<day> <Month>, <year>}
\accepted{<day> <Month>, <year>}

\begin{document}

\title{WELL: Applying Bug Detectors to Bug Localization via Weakly Supervised Learning}

\author[1]{Zhuo Li}

\author[1]{Huangzhao Zhang}

\author[1]{Zhi Jin}

\author[1]{Ge Li*}

\authormark{Zhuo Li \textsc{\emph{et al.}}}

\address[1]{\orgdiv{Key Laboratory of High Confidence Software Technologies}, \orgname{Peking University}, \orgaddress{\state{Beijing}, \country{China}}}

\corres{*Ge Li. No. 1542, No. 1 Science Building. No. 5 Yiheyuan Road, Haidian District, Beijing 100871. \email{lige@pku.edu.cn}}


\abstract[Abstract]{
Bug localization, which is used to help programmers identify the location of bugs in source code, is an essential task in software development. Researchers have already made efforts to harness the powerful deep learning (DL) techniques to automate it. However, training bug localization model is usually challenging because it requires a large quantity of data labeled with the bug's exact location, which is difficult and time-consuming to collect. By contrast, obtaining bug detection data with binary labels of whether there is a bug in the source code is much simpler. This paper proposes a WEakly supervised bug LocaLization (WELL)  method, which only uses the bug detection data with binary labels to train a bug localization model. With CodeBERT finetuned on the buggy-or-not binary labeled data, WELL can address bug localization in a weakly supervised manner. The evaluations on three method-level synthetic datasets and one file-level real-world dataset show that WELL is significantly better than the existing SOTA model in typical bug localization tasks such as variable misuse and other programming bugs.
}



\maketitle

\section{Introduction}
\label{sec:intro}

Bug localization is one of the key activities in software engineering (SE), where the practitioners are supposed to position the erroneous part of the code. Effectively automating bug localization is essential to the software developers as it can improve productivity and software quality greatly.

In the past decade, deep learning (DL) has demonstrated its great powerfulness in many SE tasks, and has achieved state-of-the-art (SOTA) performance in functionality classification \cite{Mou2016Convolutional,zhang2019novel}, code clone detection \cite{yu2019neural,wang2020clone}, method naming \cite{allamanis2016convolutional,code2vec}, code completion \cite{li2017code,liu2020self,liu2020multi} and code summarization \cite{hu2018deep,hu2018summarizing,code2seq}, \etc.
These may show the feasibility of harnessing the DL techniques to facilitate automated bug localization. Researchers have already tried to apply DL models to bug localization \cite{allamanis2018learning, vasic2019neural,vincent2020global,kanade2020learning}. GREAT \cite{vincent2020global} and CuBERT \cite{kanade2020learning} are among the SOTA DL models for bug localization and further fixing. Taking variable misuse (\Dvar) \cite{allamanis2018learning} for instance, which is one of the most thoroughly studied DL-based bug localization tasks, the DL models are supposed to locate the erroneously used variable in the given buggy code. Existing approaches, including GREAT and CuBERT, are trained in the end-to-end style, \ie, the buggy locations in the code are fine-grained annotated in the training set. 




One major challenge in existing DL solutions for bug localization is that obtaining models such as GREAT and CuBERT requires a large quantity of buggy-location-annotated training data. This kind of data provides \textit{strong supervision} as the annotations are very fine-grained and highly related to the bug localization task. 
However, such dataset with reasonable annotation quality and sufficient examples is difficult to collect or annotate, due to the huge expense of manpower and resources in real world scenarios.
According to Benton \etal \cite{benton2019defexts}, there exists only few large and publicly available bug datasets with high quality for research purpose. 
The bug localization datasets are usually obtained in two major ways -- manual annotation or automatic collection.
\ding{182} As mentioned before, it consumes a lot of manpower and resources to annotate high-quality bug localization data pairs. The annotators must be experienced software developers, and they have to take enough time to read and comprehend the code along with the bug report to locate the buggy position for every example to be annotated.
\ding{183} Automatic dataset collection, on the other hand, is much more efficient. It often utilizes web crawlers and rule-based filters to find bug fixing commits in the open source projects to locate the bugs. However, the annotation quality of automatic approaches is not guaranteed. \Eg, Lutellier \etal \cite{lutellier2020coconut} recently propose a large (million-level) program repair dataset collected from commit history of open source projects, but up to 7 of the 100 random samples are not actually bug-related commits in their manual investigations.


On the other hand, bug detection is usually a binary classification task, and utilizes the coarse-grained annotated data to train DL models. The buggy-or-not annotated data provides \textit{weak supervision} compared to bug localization, as the annotation granularity is much coarser. Data for bug detection is much easier to collect or annotate. One may automatically run tests over the projects to determine which function or file is buggy, without much effort to dive into the project nor the bug report.
Hence, in short words, bug localization data is scarce and difficult to collect or annotate, while data for bug detection is more easily accessible.

\begin{figure}
    \centering
    \includegraphics[width=0.6\columnwidth]{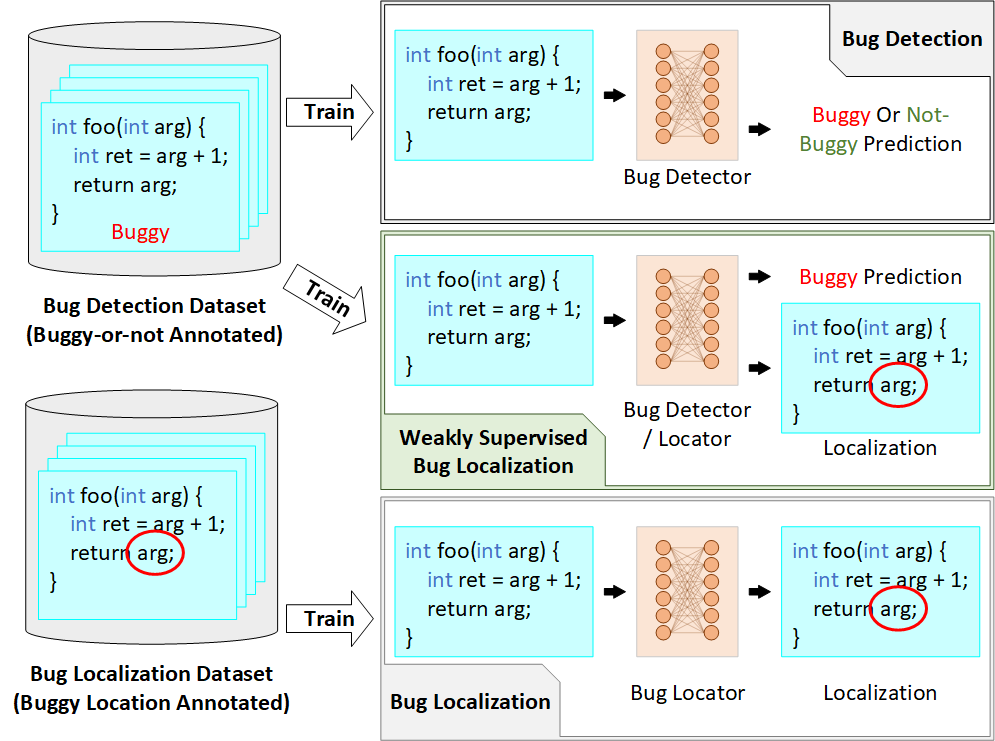}
    \caption{An demonstrative example of bug detection, bug localization, and weakly supervised bug localization. The bug detection dataset (upper left) consists of source code pieces and the corresponding binary buggy-or-not labels, which is easily accessible. While the bug localization dataset (lower left) is annotated by buggy locations in the code, which is hard to collect. Models for weakly supervised bug localization (mid right) are trained upon the detection dataset, but is able to carry out both bug detection and localization.}
    \label{fig:weak_supervision}
\end{figure}

Thence, we introduce the idea of \textit{weakly supervised learning} into bug localization. The methodology is to train the strong model (bug localization) with weak supervision signals (bug detection dataset), as illustrated in Fig. \ref{fig:weak_supervision}.
Heuristically, a bug detection model internally learns the dependency of buggy labels on the buggy portions in the code. Retrieving such knowledge embedded in the bug detection model to achieve bug localization is very desirable and feasible.
Based on such intuition, we propose \modelnameF (\modelname), which transformers the bug detection model into a bug locator without any additional trainable weights nor bug localization data.
\modelname makes full usage of the easily accessible bug detection data to tackle the lack-of-data challenge in bug localization.
To summarize the technical part, \modelname harnesses the powerfulness of the pre-trained CodeBERT model \cite{feng2020codebert}, and finetunes CodeBERT for bug localization in the weakly supervised manner. Concretely, \modelname finetunes CodeBERT on bug detection datasets, during the training stage. 
When locating bugs, \modelname acquires attention score from the finetuned CodeBERT and extracts the critical part from the input source code based on the score.
By intuition, if CodeBERT classifies a piece of code as buggy, the buggy fragment is likely to be included in the key portion of the input, which draws the model's most attention. In this way, the weakly supervised bug localization is achieved in \modelname.

At last, to demonstrate the effectiveness and capacity of \modelname, we carry out in-depth evaluations on three different synthetic token-level bug localization datasets and a real-world bug localization dataset of student programs.
The three synthetic datasets include \Dvar, bi-operator misuse (\Dbiop) and boundary condition error (\Dbound). They are the most studied synthetic bug localization datasets using DL approaches.
The student program bug localization dataset is called \Dstu. On this dataset, we train our detection model with binary test result labels (``passed'' or ``wrong answer'') and localize the buggy lines without any other annotations.
For the three synthetic datasets, on average, \modelname correctly detects and accurately locates 79.74\% of bugs, and the extended version (\modelnameE) even locates 87.57\% of bugs. Specifically, \modelname improves the localization accuracy of \Dvar to 92.28\% by over 4\% compared with CuBERT.
For \Dstu, \modelname can locate at least 1 bug for over 29\%/85\% programs when reporting the top-1/top-10 suspicious line(s) per program, outperforming the baseline models significantly.
Ablation study further demonstrates that it is feasible to apply weak supervision to other backbones, such as LSTM. The code of this project is open-sourced on Github \footnote{\url{https://github.com/Lizhmq/CodeBertRationalizer}}.

The contributions of this paper are summarized as follows: 

\begin{itemize}[leftmargin=*]
    
    \item We introduce the methodology of weak supervision into bug localization, by utilizing bug detection data. This methodology tackles the lack-of-data problem and makes full usage of the easily accessible data.
    
    
    \item We propose \modelname, which turns bug detectors into bug locators without training data of localization nor additional trainable parameters. \modelname learns to locate bugs with only bug detection datasets.
    
    
    \item We carry out in-depth evaluations to demonstrate the effectiveness of our proposed \modelname against existing SOTA DL methods for bug localization.
    Compared to the baseline models jointly obtained with strong supervision signals, \modelname trained with weak supervision produces competitive or even better performance.

    \item We demonstrate the capability of applying weak supervision to LSTM backbone through ablation study, suggesting the capacity and portability of the methodology.

\end{itemize}

\section{Related Work}
\label{sec:related}

In this section, we discuss the most relevant work to this paper, including the subject tasks of DL for bug detection and bug localization (Sec. \ref{sec:related:bug}), the methodology of weakly supervised learning (Sec. \ref{sec:related:weak}) and the techniques for DL model visualization and explanation (Sec. \ref{sec:related:rationale}).

\subsection{DL for Bug Detection and Localization}
\label{sec:related:bug}

By far, quite a lot of efforts have been made in source code processing by adopting DL techniques in the SE community \cite{zhang2019novel,wang2020clone,code2vec,code2seq,liu2020self,liu2020multi}.
To leverage DL for bug detection, Wang \etal \cite{wang2016defect} propose AST-based deep belief network for defect prediction.
Choi \etal \cite{choi2017overruns} utilize memory neural network to predict buffer overrun.
Li \etal \cite{li2018vulner} propose VulDeePecker to detect several types of vulnerabilities in source code.
Pradel and Sen \cite{pradel2018deepbugs} propose DeepBugs for bugs in function call statements and binary expressions, with a feed-forward network taking variable types and names as inputs.

As for DL-based bug localization, studies are conducted mostly on artificial synthetic datasets, where certain types of bugs are injected into the clean code to formulate buggy-location-annotated data pairs, due to the aforementioned lack-of-data problem.
Allamanis \etal \cite{allamanis2018learning} first propose the \Dvar task, which is one of the most thoroughly studied tasks in DL-based bug localization at present.
Vasic \etal \cite{vasic2019neural} employ the sequence-to-pointer (Seq2Ptr) architecture to detect, locate and fix \Dvar bugs jointly.
More recently, Hellendoorn \etal \cite{vincent2020global} propose two architectures to generate distributed representations for source code in order to locate bugs, namely Graph-Sandwich and GREAT.
Kanade \etal \cite{kanade2020learning} propose the pre-trained CuBERT for \Dvar and multiple other bug detection and localization tasks.

Although some of the aforementioned existing work train bug detection and localization model jointly \cite{vasic2019neural,vincent2020global,kanade2020learning}, \ie, the DL models are trained to detect and locate bugs simultaneously upon bug localization datasets, they do not seek to facilitate bug localization via the weak supervision signals from the bug detection (binary classification) data. As a result, the lack-of-data problem is often challenging and inevitable in traditional DL for bug localization in the real world.
In this paper, on the contrary, \modelname adopts the methodology of weakly supervised learning, and leverages the easily accessible and abundant buggy-or-not data to finetune CodeBERT as a bug detector for token-level fine-grained bug localization.

\subsection{Weakly Supervised Learning}
\label{sec:related:weak}

Weak supervision can be categorized as incomplete supervision, inexact supervision and inaccurate supervision. In this paper, we focus on inexact supervision, where the annotation of the training data is only coarse-grained labels (\eg, buggy-or-not data provides weak supervision to the bug localization task). Please refer to the survey \cite{zzh2017weak} for more detailed discussions of other types of weak supervision.

The methodology of weakly supervised learning has been demonstrated to be valid in many DL tasks.
In computer vision (CV), researchers facilitate image semantic segmentation via course-grained annotated classification datasets. The weak annotations include bounding boxes \cite{dai2015boxsup,papandreou2015weakly}, scribbles \cite{lin2016scribblesup}, and points \cite{bearman2016whats}, \etc. More recently, pixel-level segmentation by image-level annotation has been achieved through the technique of CAM \cite{gap,cam,gradcam,wei2018revisiting}.
As for natural language processing (NLP), researchers also leverage weak supervision in sequence labeling tasks, such as named entity recognition (NER), to ease the burden of data annotation \cite{ni2017weakly,patra2019weakly,lison2020named,safranchik2020weakly}.
The most relevant approach to \modelname is Token Tagger \cite{patra2019weakly}, which employs attention-based architecture for weakly supervised NER with sentence-level annotations. Token Tagger is trained to classify whether a named entity is in the sentence, and during NER, it selects the most important tokens based on the attention score.

In this paper, we adopt the idea of weakly supervised learning to train bug locators with buggy-or-not annotated data. The inner logic to leverage attention for weakly supervised learning of \modelname is inspired by Token Tagger.

\subsection{DL Model Visualization \& Explanation}
\label{sec:related:rationale}

Interpreting the DL models could aid the researchers to understand and explain the inner mechanism of neural networks, and such techniques may technically promote weakly supervised learning. In CV, at present, the CAM family \cite{gap,cam,gradcam,wei2018revisiting} are the most widely-applied and mature techniques to visualize and explain the image classifiers. It generates a heat map, where the value reflects the contribution and the importance of the corresponding pixel to the final prediction.

As for NLP, which is more relevant to our subject tasks, selective rationalization \cite{lei2016rationale,bastings2019rationale,yu2019complement,deyoung2020eraser,Jain2020bert} is one of the most effective techniques to explain sentence classifiers. It identifies the rationale, which is a subsequence of the input sentence, to best explain or support the prediction from the DL model.
The recent proposed FRESH \cite{Jain2020bert} utilizes a two-model framework for rationalization. It generates heuristic feature scores (\eg, attention score) from the subject BERT to derive pseudo binary tags on words, and finetunes another BERT in a sequence labeling manner as the rationale tagger. This work demonstrates the powerfulness of pre-trained models in comprehending and explaining the NLP models.

Inspired by FRESH, we employ a simplified framework in \modelname, which utilizes one single finetuned CodeBERT \cite{feng2020codebert} bug detector to generate attention scores for bug localization. As demonstrated in selective rationalization, attention in CodeBERT is supposed to mine the portions of the code informative for bug detection, and such portions are supposed to be bugs.

\section{Problem Definition \& Preliminary}
\label{sec:prelim}

\begin{table}[t]
    \centering
    \caption{Summary of notations and symbols in this paper.}
    \label{tab:notation}
    \begin{tabular}{cl}
    \toprule
Notation & Definition \\
    \midrule
$\mathcal{D}^t,\mathcal{D}^v,\mathcal{D}^e$ & Dataset for training, validation \& evaluation. \\
$\mathcal{X},\mathcal{Y}$ & Source code space \& annotation space. \\
$x=(t_1,\cdots,t_l)$ & Token sequence of the source code. \\
$(s_1,\cdots,s_{l'})$ & Subtoken sequence of the source code. \\
$C,\Theta_C$ & DL model \& its trainable parameters. \\
$y, \tilde y$ & Annotation \& prediction from the model. \\
$\mathcal{L}(\tilde y, y)$ & Loss function. \\
$Q,K,V$ & Query, key \& value matrices in attention. \\
$h=(h_0,\cdots,h_{l'})$ & Context-aware hidden states. \\
$\alpha=(\alpha_1,\cdots,\alpha_{n_h})$ & Multi-head attention with $n_h$ heads. \\
    \bottomrule
    \end{tabular}
\end{table}

In this section, we first provide our formal definition of the object tasks (Sec. \ref{sec:prelim:bugloc}). Then we explain the multi-head attention mechanism in the transformer model (Sec. \ref{sec:prelim:transformer}), based on which we facilitate \modelname. And at last, we introduce the large CodeBERT model pre-trained for source code (Sec. \ref{sec:prelim:codebert}), which acts as the backbone of \modelname. The symbols we employ in this paper are summarized in Table \ref{tab:notation}.

\subsection{Bug Detection \& Localization}
\label{sec:prelim:bugloc}

\paragraph{Dataset} Both bug detection and localization are supervised tasks, since the datasets are annotated. A typical dataset consists of multiple pairs of examples, \ie, $\mathcal{D}=\{(x_1,y_1),\cdots,(x_n,y_n)\}$, where $n$ refers to the size of the dataset. A pair of example $(x,y)\in\mathcal{D}$ includes a source code piece $x\in\mathcal{X}$ and its corresponding annotation $y\in\mathcal{Y}$. The annotation space varies for different tasks, which will be defined later in this section. The source code is already tokenized, \ie, $x=(t_1,\cdots,t_l)$, where $t_i$ is the $i$-th token in $x$ and $l$ is the length of the token sequence.

\paragraph{Model} A DL model $C$ takes the tokenized source code $x$ as input, and outputs $\tilde{y}\in\mathcal{Y}$ as the prediction. Note that the output format also varies for different tasks. $C$ is supposed to produce $\tilde y=y$, where the prediction exactly matches the ground-truth. To obtain the optimal parameters in $C$ is to optimize the objective function $\min_{\Theta_C} \sum_{(x,y)\in\mathcal{D}^t}\mathcal{L}(\tilde y,y)$, where $\mathcal{L}(\cdot)$ is the loss function (usually cross entropy), which measures the similarity of $\tilde y$ and $y$.
For those $C$'s which require different input formats (\eg, AST or graph), we assume that they process the format internally.

\paragraph{Bug detection} We define bug detection in method-level. Bug detection is to determine whether a bug exists in the given method or function. Therefore, the annotation space for bug detection is binary, \ie, $\mathcal{Y}=\{0,1\}$, where $y=1$ suggests that the corresponding function $x$ is buggy, while $y=0$ the opposite.
One often adopted approach to build a bug detector is to first generate an encoding vector of the source code via a code representation model, and then classify the encoding with a feed forward network.

\paragraph{Bug localization} Bug localization is to determine which token subsequences cause a bug in the given code. The annotation for bug localization is defined as a sequence of binary labels, \ie, $\mathcal{Y}=\{0,1\}^l$. Each token annotation $y_i$ in $y=(y_1,\cdots,y_l)$ is binary, where $y_i=1$ suggests that $t_i$ participates in the bug, while $y_i=0$ the opposite.
Hence, bug localization can be viewed as a sequence tagging problem on source code.
Bug locators take $x$ as input and binarily tags each $t_i$.

In this paper, as a very early step in this area, we focus on the fine-grained synthesized bugs, where the bug is caused by one or two tokens, \eg, \Dvar by a single variable misuse, \Dbiop by a single bi-operator misuse, and \Dbound by a single inequality operator misuse. We propose and evaluate \modelname on these datasets. The proposed weakly supervised \modelname has potential to be extended to detect and locate other more complex bugs. We leave it for future work.

\paragraph{Existing strongly supervised bug localization} Existing approaches are mainly based on the Seq2Ptr framework \cite{vasic2019neural,vincent2020global,kanade2020learning}. The DL model is trained on the buggy-location-annotated datasets with strong supervision. Concretely, the model encodes the source code, and computes attentions based on the token representations. The ``pointer'' points to the token with the highest attention score as the predicted buggy location, and for non-buggy (clean) code, the ``pointer'' points to a special ``clean'' token inserted in the sequence.

\subsection{Attention in Transformer}
\label{sec:prelim:transformer}
         
The attention technique is first proposed in Seq2Seq model \cite{bahdanau2015attention}, which aims to selectively focus on parts of the source sequence during prediction.
The attention function maps multiple queries and a set of key-value pairs to a weighted sum of the values, where the weight assigned to each value is computed with the query and the corresponding key \cite{vaswani2017transformer}.

\paragraph{Query, key \& value} The queries, the keys and the values are all vectors. The queries $Q=(Q_1,\cdots,Q_n)$ are the points of interest, the values $V=(V_1,\cdots,V_l)$ are the vectors to be weighted, and the keys $K=(K_1,\cdots,K_l)$ are the ``descriptions'' of $V$. Note that each $K_i$ and $V_i$ are paired. We query $Q_i$ against the key $K_j$ about how much portion of the value $V_j$ should be included in the weighted-sum output.

\paragraph{Scaled self-attention} The scaled self-attention measures how relevant the tokens in the given sequence $x$ are to each other. Therefore, $Q$, $K$ and $V$ are all generated from $X$, where $X\in\mathbb{R}^{l\times d}$ is the embedding matrix of $x$ and $d$ is the dimension of the embedding space. Concretely, $Q=W_QX$, $K=W_KX$, and $V=W_VX$, where $W_Q$, $W_K$ and $W_V$ are trainable parameters. Scaled self-attention is computed as Eq. \ref{eq:transformer:attn}, where $\alpha(Q,K)$ produces the probabilistic attention score, and $\sigma$ is the softmax function. 
$\alpha_{ij}$ (the element of the $i$-th row and the $j$-th column in $\alpha$) is the attention score of query $Q_i$ against key $K_j$. $\alpha_{ij}$ reflects how relevant $t_i$ and $t_j$ are in $x$. For other types of attention, please refer to \cite{luong2015effective}.

\begin{equation}
    {\rm Attention}(Q,K,V)=\alpha(Q,K)V=\sigma\left(\frac{QK^T}{\sqrt{d}}\right)V \label{eq:transformer:attn}
\end{equation}





\paragraph{Multi-head attention} The multi-head attention mechanism is proposed in transformer \cite{vaswani2017transformer} to focus on different types of information in the sequence.
It consists of multiple sets of scaled self-attention, and each attention is called a head, where the parameters in each head are not shared. The outputs of these heads are concatenated and linearly transformed to integrate the information and generate the final representation. The overall formulation of multi-head attention in the transformer architecture is deducted as below:

\begin{equation}
\begin{aligned}
    {\rm MultiHead}(X)&={\rm Concat}({\rm head}_1,\cdots,{\rm head}_{n_h})W_O,\\ 
    {\rm~where~}{\rm head}_i&={\rm Attention}_i(Q_i,K_i,V_i),
\end{aligned}
\end{equation}

\noindent where $n_h$ is the head number, and $W_O$, along with $W_{Qi}$, $W_{Ki}$, $W_{Vi}$ in the $i$-th head, are trainable parameters.


\paragraph{Transformer layer}
The transformer architecture \cite{vaswani2017transformer} is composed of a stack of identical transformer layers.
Each layer consists of a multi-head attention and a fully connected neural network. These two modules are linked through residual connection \cite{resnet} and additional layer normalization \cite{layer_norm}. The $i$-th layer takes $X_i$ as input and computes $X_{i+1}$ for the $i+1$-th layer ($X_1=X$). Please refer to ``Layer 1'' in Fig. \ref{fig:cbl} for the detailed structure of the transformer layer.

\subsection{Pre-trained Models}
\label{sec:prelim:codebert}

With the rapid increase of the amount of accessible training data and computing power, the ways of using DL models have also changed greatly. Recently, researchers have demonstrated the SOTA performance of large pre-trained models on various tasks across different domains including CV and NLP. Compared to traditional training from scratch, pre-training on general tasks and fine-tuning on specific downstream tasks result in a significant performance improvement. The idea of pre-training is introduced to the field of SE lately, resulting in the CodeBERT \cite{feng2020codebert} model pre-trained for source code.

\begin{figure}[t]
    \centering
    \includegraphics[width=0.6\linewidth]{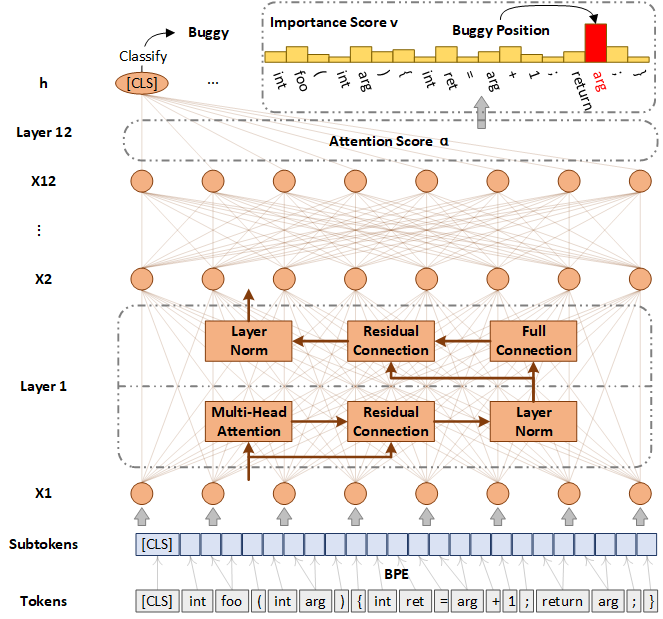}
    \caption{An illustrative example of \modelname based on the CodeBERT backbone. We plot the computation flow chart of transformer layer only for ``Layer 1'' to save space. The top-left part is for bug detection, while the top right part is for bug localization.}
    \label{fig:cbl}
\end{figure}

\paragraph{CodeBERT}
Following the work of BERT \cite{bert} and RoBERTa \cite{roberta} in NLP, Feng \etal. \cite{feng2020codebert} propose CodeBERT for source code, which employs the identical architecture and model size with RoBERTa. The general architecture of CodeBERT is shown in Fig. \ref{fig:cbl}.
Specifically, CodeBERT consists of 12 transformer layers (Layer 1 to 12). The hidden size, attention head number and feed-forward size in each layer are 768, 12 and 3,072 respectively, leading to overall 125 million parameter size. CodeBERT is pre-trained on the CodeSearchNet dataset \cite{codesearchnet} with more than 8 million functions across six programming languages (\ie, Python, Java, JavaScript, Php, Ruby, and Go). After finetuning, CodeBERT is capable of performing code classification and code generation (as an encoder) tasks, and producing SOTA results in various SE tasks \cite{codexglue}.

\section{\modelname - Weakly Supervised bug localization}
\label{sec:method}

In this section, we illustrate the proposed \modelname in detail. We first give a high-level overview of \modelname in Sec. \ref{sec:method:overview}, which locates bugs through learning upon detection tasks in a weakly supervised manner. Then, we present details of training the model and locating bugs in Sec. \ref{sec:method:detect} and Sec. \ref{sec:method:attention}. Finally, we extend \modelname by leveraging a small amount of buggy-location-annotated data for validation when available.


\subsection{Overview of \modelname}
\label{sec:method:overview}

We first present the high-level idea of our proposed \modelname, utilizing the CodeBERT model as the backbone to locate bugs in source code in a weakly supervised manner. It takes three steps to achieve weakly supervised learning:
\ding{182} \textit{finetuning} CodeBERT on bug detection datasets,
\ding{183} \textit{predicting} whether a piece of suspicious code is buggy or not, and 
\ding{184} \textit{locating} the buggy position based on the attention score. One may note that finetuning (or training) and predicting constitute the common learning paradigm of classification tasks such as bug detection. We follow this paradigm to obtain CodeBERT as bug detectors, and facilitate bug localization according to the attention score. Intuitively, if CodeBERT is capable to detect bugs, the code segment that draws the most attention should be related to the bug. Therefore, the multi-head attention should reveal why and how CodeBERT detects bugs, and in this way, we achieve weakly supervised bug localization.


\modelname takes the source code token sequence as input, and outputs the predicted buggy-or-not label and the bug location, as illustrated in Fig. \ref{fig:cbl}.
Specifically, during localization, \modelname first carries out binary classification upon the code with the finetuned CodeBERT (``Tokens'' up to ``$h$'' in Fig. \ref{fig:cbl}). If the prediction is negative, \modelname terminates, as the code is considered as clean (not buggy); otherwise, the algorithm continues. During prediction, the multi-head attention scores of the last layer ($\alpha$) are also retrieved. \modelname aggregates the multi-head attention scores to compute the token-level importance score $v=(v_1,\cdots,v_l)$, where $v_i$ suggests the likeliness of the token $t_i$ in the code $x$ to cause the bug.


\subsection{Learning to Detect Bugs}
\label{sec:method:detect}

As aforementioned, \modelname finetunes the CodeBERT model upon bug detection datasets, which provide weak supervision for bug localization. As binary labeled bug detection datasets are easily accessible, CodeBERT is completely capable to learn the buggy patterns in the source code. This section illustrates the finetuning and predicting steps.


\paragraph{Forward computation} Forward computation means prediction in DL. According to the definition of bug detection in Sec. \ref{sec:prelim:bugloc}, CodeBERT takes the source code $x = (t_1, ..., t_l)$ as input and outputs $\tilde{y}$ as the predicted label. The forward computing process is demonstrated in Fig. \ref{fig:cbl}. Concretely, given the source code token sequence $x$, we split them into subtokens $s_1, s_2, ..., s_{l'}$ using BPE \cite{bpe}, where $l'$ is the length of the subtoken sequence. Special token $s_0={\rm [CLS]}$ for classification is also inserted at the very beginning of the subtoken sequence. Then, the 12 transformer layers generates the context-aware vector representation $h=h_0,...,h_{l'}$ of subtokens.
We preserve only $h_0$ as the aggregated representation of the whole piece of code, and discard other $h_i$'s. Finally, we feed $h_0$ into a feed-forward neural network with softmax output, producing the predicted probabilities $p = (p_0, p_1)^T$ of the two classes ($p_1$ for buggy and $p_0$ for clean). The binary prediction $\tilde y$ is made based on $p_1$, \ie, $\tilde y=1$ if $p_1\ge0.5$, otherwise, $\tilde y=0$.



\paragraph{Training objective} We finetune CodeBERT on the training set of bug detection via back propagation and gradient descent. The loss function is the commonly adopted cross-entropy loss for classification, which is formulated as below, where $\Theta_C$ refers to the weights in the CodeBERT bug detection model:

\begin{equation}
    \min_{\Theta_C}\mathcal{L} = \mathbb{E}_{(x,y)\in \mathcal{D}^t}\left(-y\mathop{log}p_1-(1-y)\mathop{log}p_0\right).
\end{equation}

\subsection{Localization via Multi-head Attention}
\label{sec:method:attention}

Empirically, the CodeBERT bug detector obtained in Sec. \ref{sec:method:detect} makes prediction based on some certain features embedded within the code. Such features, which lead to buggy prediction, are likely to be related to the bugs. On the other hand, the multi-head attention provides which parts of the source code the CodeBERT model focuses on. By analyzing the multi-head attention, we may locate bugs via the CodeBERT bug detector. The algorithm is presented in Algo. \ref{algo:codebert-l}.


\paragraph{Forward computation} In order to determine whether the input code piece is buggy or not, and to retrieve the attention scores $\alpha$ for further localization procedures, a forward computation (Line \ref{algo:bpe} to \ref{algo:argmax1} in Algo. \ref{algo:codebert-l}) is necessary.
After retrieving $\alpha$, \modelname performs two aggregations (multi-head and subtoken) to obtain the token-level importance score $v$.


\paragraph{Aggregation of multiple heads} The attention score $\alpha$ (as a 3-dimensional tensor) includes all attention heads from the last transformer layer in CodeBERT. The $i$-th attention head is denoted as $\alpha_i$, in which the $j$-th row ($\alpha_{ij}$) suggests the probabilistic attention of $s_j$ to all other subtokens of the code. Specifically, $\alpha_{i0}$ suggests the importance of each subtoken for the classification from the $i$-th head. 
In order to take all heads into consideration, we adopt average aggregation of all the $n_h$ heads as Line \ref{algo:ave} in Algo. \ref{algo:codebert-l}:

\begin{equation}
    \alpha' = {\rm AGG\_HEAD}(\alpha) = \sum\nolimits_{i=1}^{n_h}\frac{\alpha_{i0}}{n_h}, \label{eq:ave}
\end{equation}

\noindent where the output $\alpha'$ is a sequence of probabilities suggesting the importance of each corresponding subtoken.


\begin{algorithm}[t]
    \caption{Localization algorithm by \modelname.}
    \label{algo:codebert-l}
    \begin{algorithmic}[1]
        \Require Source code $x = t_1, t_2, ..., t_l$
        \Ensure Predicted label $y$ and buggy fragment $x_{buggy}$.
        \State $s_1, s_2, ..., s_{l'} \leftarrow BPE(t_1, t_2, ..., t_l)$ \label{algo:bpe} \Comment{Tokenize the input code}
        \State $h, \alpha \leftarrow {\rm CodeBERT}([CLS], s_1, s_2, ..., s_{l'})$ \label{algo:encode}
        \State $p \leftarrow {\rm CLASSIFY}(h_0)$ \Comment{Obtain prediction with CodeBERT}
        \State $\tilde{y} \leftarrow \mathop{argmax}_ip_i$ \label{algo:argmax1}
        \If{$\tilde{y} = 0$} \Comment{Source code classified as non-buggy} \\ 
            \quad \  \Return CLEAN, NONE \label{algo:negative}
        \Else
            \State $\alpha' \leftarrow
            {\rm AGG\_HEAD}(\alpha)$
            \label{algo:ave} \Comment{Aggregate importance scores from different attention heads}
            \State $a, b \leftarrow {\rm ALIGN}(x, s)$ \label{algo:align}
            \Comment{Align subtokens with code tokens}
                \State $v \leftarrow {\rm AGG\_SUBTOKEN}(a,b,\alpha')$ \Comment{Sum subtokens' importance score}
                \label{algo:sum}
            \State $k \leftarrow \arg\max_{i}\sum_{j=i}^{i+N-1}v_j$ \label{algo:loc} \Comment{Select the position with the highest importance score}  \\
            \quad \  \Return BUGGY, $(t_k,\cdots,t_{k+N-1})$
        \EndIf
    \end{algorithmic}
\end{algorithm}

\paragraph{Alignment of subtokens} Due to BPE, a token $t_i$ may be splitted into multiple subtokens, and the concatenation of these subtokens makes the original $t_i$ \ie, $t_i={\rm CONCAT}(s_{a_i},s_{a_i+1},\cdots,s_{b_i})$,
where $a_i$ and $b_i$ refer to the beginning and ending indices in the subtoken sequence of the token $t_i$ respectively. $a$ and $b$, which are vectors constituted by $a_i$ and $b_i$, are retrieved by aligning the token sequence $x$ and the subtoken sequence $s$ as Line \ref{algo:align} in Algo. \ref{algo:codebert-l}.
Concretely, BPE in CodeBERT tags the subtokens with a special character ``$\dot{G}$'', referring to the beginning of a token. Therefore, we perform a scanning over $x$ and $s$ to collect $a_i$ and $b_i$ for each $t_i\in x$.

\paragraph{Aggregation of subtokens}
Since $\alpha'$ refers to the subtoken-level importance score while we require the token-level importance score, an aggregation upon the subtokens is necessary.
After alignment, \modelname carries out additive aggregation to map the attention from the subtokens to the corresponding tokens (Line \ref{algo:sum} in Algo \ref{algo:codebert-l}), resulting in the importance score $v=(v_1,\cdots,v_l)$. Each $v_i$ is computed as $v_i=\sum\nolimits_{j=a_i}^{b_i}\alpha'_j.$



\paragraph{Localization} The importance score $v_i$ suggests how informative $t_i$ is to the buggy prediction of the CodeBERT bug detector. In other words, those tokens with high importance scores are more likely to be related to the bug. As the bug localization task can be treated as a sequence tagging problem over the source code (Sec. \ref{sec:prelim:bugloc}), \modelname assumes the buggy fragment to be a consecutive token subsequence of length $N$, where $N$ is a hyper-parameter.
With this assumption, \modelname utilizes a slide window of size $N$ to compute the importance score of all fragments, and selects the one with the largest score as the buggy fragment (Line \ref{algo:loc} in Algo. \ref{algo:codebert-l}). The buggy fragment $x_{buggy}=(x_k,\cdots,x_{k+N-1})$ is selected as:


\begin{equation}
    k = \arg\max_{i}\sum\nolimits_{j=i}^{i+N-1}v_j. \label{eq:argmax_locate}
\end{equation}

Please note the two assumptions in Eq. \ref{eq:argmax_locate}: \ding{182} The buggy tokens are no more than $N$, and \ding{183} the buggy tokens are consecutive. When the assumptions are not satisfied, we could expand \modelname by adopting the threshold activated inconsecutive selection strategy. We leave this to the future work.


\subsection{Extension with Fine-grained Supervision}
\label{sec:method:extension}

So far, we have illustrated all details in \modelname, including the forward computation of the CodeBERT backbone, the finetuning protocol upon buggy-or-not datasets, and the localization process. In other words, \modelname does not require nor rely on any buggy location annotations for training. However, in the real world, although the fine-grained well-annotated examples are hard to collect, we can still obtain a small amount of them by all means. We introduce the extended \modelname, namely \modelnameL{$k$}, in this section, which leverages these fine-grained buggy-location-annotated examples for validation.


Even though multi-head attention is designed to focus on different features in the input sequence \cite{vaswani2017transformer}, recent researches have shown that only a small subset of the heads are specialized for the downstream task, while the other heads are dispensable and can even be pruned without losing much performance \cite{head_pruning,head_pruning2}. Therefore, those unimportant heads in \modelname may have negative impacts after the average aggregation, which we indeed encounter during our experiments. On the other hand, aggregation of only the important heads is supposed to be competitive or even better. Therefore, \modelnameL{$k$} is proposed as an extension of \modelname, which selects and aggregates only the top-$k$ important attention heads.


To measure the importance of each attention head, we utilize the fine-grained well-annotated examples as validation. Instead of direct aggregation, we evaluate the bug localization performance of the $i$-th attention head against the validation set (buggy-location-annotated), by setting $\alpha'=\alpha_{i0}$, creating \modelnameL{H$_i$}.
The performance of \modelnameL{H$_i$} is considered as the importance of the $i$-th head.
Then, we aggregate only the top-$k$ important heads, resulting in \modelnameL{$k$}. Note that \modelnameE refers to no attention head aggregation at all (using only the most important head), and \modelnameL{$n_h$} refers to \modelname itself (aggregation of all heads). During our experiments, we evaluate \modelnameE, considering the extreme case.



\section{Evaluation}
\label{sec:evaluation}

\begin{table}[t]
	\centering 
	\caption{Information of the subject tasks \& datasets}
	\label{tab:dataset}
	    \begin{threeparttable}
	        \begin{tabular}{lcccccc}
			\toprule
Dataset & Language & Train \# &  Valid \# & Test \# & $\bar l$\tnote{*} & $\Psi$\tnote{*} \\
			\midrule
\Dvar & Python2 & $\sim$1.6M & $\sim$170k & $\sim$886k & 73 & \ding{51} \\
\Dbiop & Python2 & $\sim$460k & $\sim$49k & $\sim$250k & 123 & \ding{51} \\
\Dbound & Java & $\sim$180k & $\sim$26k & $\sim$52k & 146 & \ding{51} \\
\hline
\Dstu & C & $\sim$235k & $\sim$26k & $\sim$17k & 170 & \ding{51} \\
			\bottomrule
		\end{tabular}
		\begin{tablenotes}\footnotesize
		        \item[*] $\bar l$ = ``Average length'', $\Psi$ = ``Well balanced''.
		\end{tablenotes}
	    \end{threeparttable}
\end{table}

\subsection{Experimental Settings}
\label{sec:evaluation:settings}

\paragraph{Subject tasks \& datasets}
We evaluate \modelname, against other approaches, with three synthetic bug detection/localization tasks, \ie, \Dvar \cite{allamanis2018learning}, \Dbiop \cite{kanade2020learning}, \Dbound, and a student program bug localization task (\Dstu).
\Dvar \cite{allamanis2018learning} is a variable misuse detection and localization dataset of Python2 functions. The task is to detect whether there is a misused variable name in each function and locate it.
\Dbiop is an operator misuse detection benchmark of Python2 functions proposed by Kanade et al. \cite{kanade2020learning}. Bugs are introduced by substituting one bi-operator with a wrong but type-compatible random one (\eg, ``+''$\Leftrightarrow$``-'', ``*''$\Leftrightarrow$``/'', ``is''$\Leftrightarrow$``is not''). We extend this task to further bug localization by requiring the models to locate the substituted operator.
\Dbound is a boundary condition error detection and localization dataset of Java methods proposed in this paper. The off-by-one bug is brought into Java methods by adding/remove the equal condition in binary comparison operators. The details of obtaining this dataset are described in the Appendix.
The bug types in the three synthetic datasets are different, but all of them are caused by one or two tokens (\eg ``is not''$\Leftrightarrow$``is'' in \Dbiop).

\Dstu \cite{gupta19attr} is a bug localization dataset of student programs written in the C language. The dataset is collected from 28,331 student submissions for 29 programming tasks with 231 test cases in total. During training, models learn to predict whether a program can pass a test case, \ie, the input is a (program, test case) pair. For localization evaluation, models need to locate the buggy lines in the programs in the test set. Please refer to the original paper for more details.
The statistical and other information about the datasets is listed in Table \ref{tab:dataset}.

\paragraph{Baseline Models} 
For the synthetic datasets, SOTA DL-based solutions for bug localization are mostly Seq2Ptr architecture \cite{vasic2019neural}. Seq2Ptr employs the DL encoders to compute attention queried by a trainable vector upon the input code piece, and the pointer to the buggy position is generated from the attention ($\arg\max$). Different from our proposed \modelname, Seq2Ptr models are trained with the bug localization dataset with strong buggy-location-annotated supervision signals. We employ previously proposed SOTA models as the baseline, including GREAT \cite{vincent2020global} and CuBERT \cite{kanade2020learning}. For experiments on \Dstu, we employ NeuralBugLocator, two SOTA program-spectrum based and one syntactic difference based bug localization baselines \cite{gupta19attr}.
The implementation details (including hyper-parameters) are listed in the Appendix.

\paragraph{Evaluation metrics}
For synthetic datasets, we employ \textit{classification accuracy} (Acc$_D$), \textit{precision} (P$_D$), and \textit{recall} (R$_D$) as the evaluation metrics for bug detection and \textit{localization accuracy} (Acc$_L$) for bug localization.
For \Dstu, we employ top-$k$ ($k=1,5,10$) localization accuracy, which represents the percentage of programs that at least one buggy line is located successfully when top-$k$ suspicious lines are reported. We evaluate the accuracy on 1,449 programs in the test set as Gupta et al. \cite{gupta19attr} do.

\subsection{Evaluation on Synthetic Datasets}
\label{sec:experiments:RQ1}

\begin{table}[t] 
  \begin{minipage}[t]{0.52\textwidth} 
    \centering
        	\begin{threeparttable}
        	
    	\caption{Bug detection/localization results on \Dvar.}
        \label{tab:res_var}
        		\begin{tabular}{cccccccc}
        			\toprule
        \multicolumn{2}{c}{Sup. sig.} & \multirow{2}*{Model} & \multicolumn{3}{c}{Detection} && Localization \\
            \cmidrule{1-2}
            \cmidrule{4-6}
            \cmidrule{8-8}
        D\tnote{\textdagger} & L\tnote{\textdagger} & & P (\%) & R (\%) & Acc. (\%) && Acc. (\%) \\
        			\midrule
        \ding{51} & \ding{51} & GREAT & 91.38 & 91.69 & 89.91 && 85.53 \\
        \ding{51} & \ding{51} & CuBERT & 93.53 & 91.76 & 92.69 && 88.22 \\
        			\midrule
        \ding{51} & \ding{55} & \modelname & {\bf 94.34} & {\bf 96.12} & {\bf 95.20} && {\bf 92.28} \\
        \ding{51} & \ding{55} & \modelnameE & {\bf 94.34} & {\bf 96.12} & {\bf 95.20} && 92.08 \\
        \bottomrule
        		\end{tabular}
        		\begin{tablenotes}\scriptsize
        \item[\textdagger] D = ``Detection supervision'', L = ``Localization supervision''.
        		\end{tablenotes}
        	\end{threeparttable}
  \end{minipage}%
  \begin{minipage}[t]{0.52\textwidth} 
    \centering
        \begin{threeparttable}
            	\caption{Bug detection/localization results on \Dbiop}
	    \label{tab:res_biop}
        		\begin{tabular}{cccccccc}
        			\toprule
        \multicolumn{2}{c}{Sup. sig.} & \multirow{2}*{Model} & \multicolumn{3}{c}{Detection} && Localization \\
            \cmidrule{1-2}
            \cmidrule{4-6}
            \cmidrule{8-8}
        D\tnote{\textdagger} & L\tnote{\textdagger} && P (\%) & R (\%) & Acc. (\%) && Acc. (\%) \\
        			\midrule
        \ding{51} & \ding{51} & GREAT & 82.32 & 81.98 & 82.92 && 76.53 \\
        \ding{51} & \ding{55} & CuBERT & 86.64 & 88.66 & 87.49 &&  \textbf{85.14} \\
                    \midrule
        \ding{51} & \ding{55} & \modelname & {\bf 92.70} & {\bf 90.50} & {\bf 91.71} && 83.08 \\
        \ding{51} & \ding{55} & \modelnameE & {\bf 92.70} & {\bf 90.50} & {\bf 91.71} && 83.44 \\
        			\bottomrule
        		\end{tabular}
        		\begin{tablenotes}\scriptsize
        \item[\textdagger] D = ``Detection supervision'', L = ``Localization supervision''.
        		\end{tablenotes}
        	\end{threeparttable}
  \end{minipage}
\end{table}

\begin{table}[t]
	\centering
	\caption{Bug detection/localization results on \Dbound}
	\label{tab:res_bound}
	\begin{threeparttable}
		\begin{tabular}{cccccccc}
			\toprule
\multicolumn{2}{c}{Sup. sig.} & \multirow{2}*{Model} & \multicolumn{3}{c}{Detection} && Localization \\
    \cmidrule{1-2}
    \cmidrule{4-6}
    \cmidrule{8-8}
D\tnote{\textdagger} & L\tnote{\textdagger} && P (\%) & R (\%) & Acc. (\%) && Acc. (\%) \\
			\midrule
\ding{51} & \ding{51} & GREAT & 85.87 & 88.11 & 87.08 && 84.77 \\
\ding{51} & \ding{51} & CuBERT & 90.12 & 92.36 & 91.12 && {\bf 90.10} \\
            \midrule
\ding{51} & \ding{55} & \modelnameL{lstm} & 89.00 & 90.40 & 89.63 && 34.96 \\
            \midrule
\ding{51} & \ding{55} & \modelname & {\bf 91.50} & {\bf 93.00} & {\bf 93.10} && 63.87 \\
\ding{51} & \ding{55} & \modelnameE & {\bf 91.50} & {\bf 93.00} & {\bf 93.10} && 87.19 \\
			\bottomrule
		\end{tabular}
		\begin{tablenotes}\scriptsize
\item[\textdagger] D = ``Detection supervision'', L = ``Localization supervision''.
		\end{tablenotes}
	\end{threeparttable}
\end{table}

To demonstrate \modelname's effectiveness in detecting and locating bugs, we evaluate the performance of WELL on the three synthetic datasets and compare it with the baselines.

\paragraph{Bug detection}
The precision (P$_D$), recall (R$_D$), and accuracy (Acc$_D$) of \modelname and baseline models for detecting \Dvar, \Dbiop, and \Dbound bugs are listed in Table \ref{tab:res_var}-\ref{tab:res_bound}. Note that \modelnameE and \modelname share the same finetuned CodeBERT backbone, the detection performance is identical. On average, \modelname produces 92.85\% precision, 93.21\% recall, and 93.33\% accuracy in detecting the three types of bugs. The performance of \modelname on bug detection is significantly better than GREAT and CuBERT, which is attributed to the powerful CodeBERT backbone.

\begin{figure}[t]
    \centering
    \subfigure[\Dvar]{
        \includegraphics[width=0.44\columnwidth]{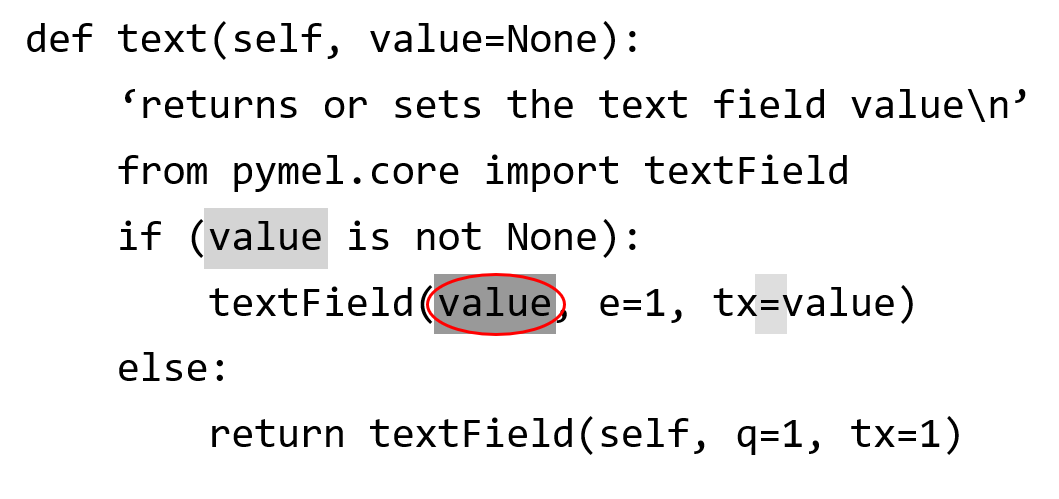}
        \label{fig:visual:var}
    }
    \subfigure[\Dbiop]{
        \includegraphics[width=0.44\columnwidth]{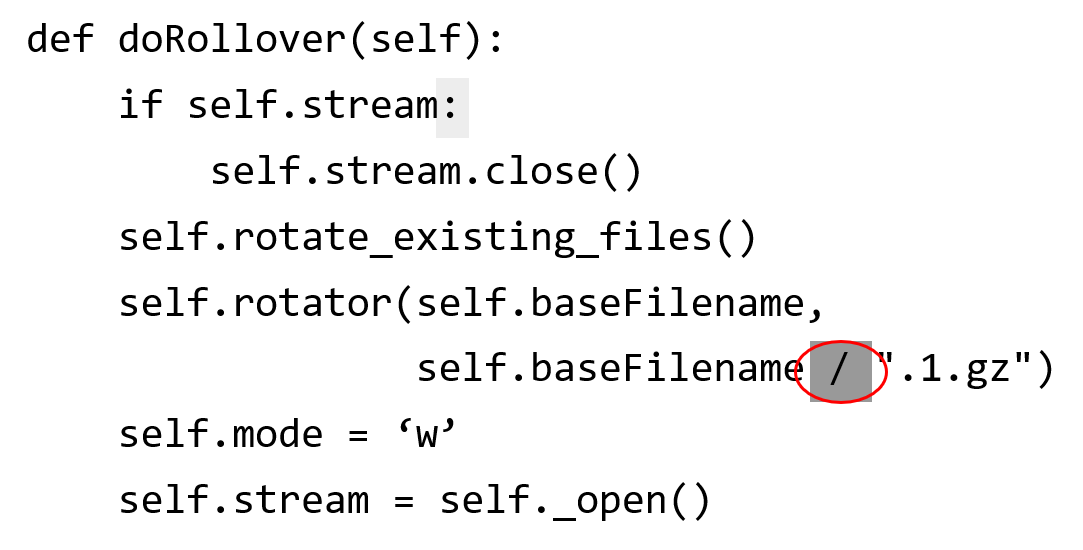}
        \label{fig:visual:op}
    }
    \caption{Visualization of the importance score produced by \modelname. All examples are correctly handled by \modelname. The red circle suggests the buggy location of ground-truth. The gray-scale of the background represents the importance score of the corresponding token by \modelname.}
    \label{fig:visual}
\end{figure}

\paragraph{Bug localization}
On average, \modelname and \modelnameE correctly locate 79.74\% and 87.57\% of bugs as shown in Table \ref{tab:res_var}-\ref{tab:res_bound}. Compared with random picking, whose accuracy is about 1\%, \modelname is able to locate the three kinds of bugs, even though it has no direct supervision signals for localization during finetuning.
To our surprise, the weakly supervised \modelname is comparable to SOTA supervised models, GREAT and CuBERT. 
Specifically, the localization accuracy of \modelname is 6.75\%  higher than GREAT in \Dvar, and 4.06\% higher than CuBERT. As for \Dbiop and \Dbound, the results of \modelnameL{1} are only ~2\% lower than CuBERT.
In addition, we notice that \modelname and \modelnameE produce similar accuracy in \Dvar and \Dbiop (with differences less than 0.5\%), but \modelnameE outperforms \modelname by 23.32\% accuracy in \Dbound. This phenomenon is investigated and discussed in Section \ref{sec:experiments:RQ3} (``Ablation on Attention Heads'') later.

\paragraph{Case study} Fig. \ref{fig:visual} shows two cases from the \Dvar and \Dbiop separately. \modelname predicts them as buggy correctly and locates the bugs accurately.
The darker background of a token ($t_i$) refers to the higher importance score ($v_i$) from \modelname of this token. The most important tokens (darkest) and the buggy locations (red circle) coincide in the figures, which, to a certain extent, shows the effectiveness of the weakly supervised \modelname.
On the other hand, important tokens with dark backgrounds are scarce and concentrated in the figures, meaning that CodeBERT in \modelname is capable of learning the relation between the buggy position in the code and the given buggy-or-not supervised signal during finetuning. This furthermore demonstrates the feasibility and effectiveness of weak supervision in bug localization. Please refer to the Appendix for more visualized cases.

The experiment results and case study indicate that \modelname is feasible and effective in detecting and locating bugs in the evaluated datasets.

\begin{table}[t]
    \centering
	\caption{Bug localization results on \Dstu.}
	\label{tab:res_stu}
	\begin{threeparttable}
		\begin{tabular}{ccccc}
			\toprule
			 \multirow{2}*{Model} & \multicolumn{3}{c}{Localization Result} \\
			  & Top-10 & Top-5 & Top-1 \\
			\midrule
			 Tarantula & 1,141 (78.74\%) & 791 (54.59\%) & 311 (21.46\%) \\
			 Ochiai & 1,151 (79.43\%) & 835 (57.63\%) & 385 (26.57\%) \\
			 Diff-based & 623 (43.00\%) & 122 (8.42\%) & 0 (0.00\%) \\
             NBL & 1,164 (80.33\%) & 833 (57.49\%) & 294 (20.29\%) \\
            \midrule
             \modelname & {\bf 1,240} (85.58\%) & {\bf 931} (64.25\%) & {\bf 421} (29.05\%) \\
            \bottomrule
		\end{tabular}
	\end{threeparttable}
\end{table}

\subsection{Evaluation on Student Programs}
\label{sec:experiments:RQ2}

We further evaluate the performance of \modelname by locating semantic bugs in student programs and compare \modelname with existing SOTA deep models NBL \cite{gupta19attr}, two program-spectrum-based methods (Tarantula \cite{abreu06spectrum}, Ochiai \cite{jones03spectrum}) and one syntactic difference based technique. We evaluate the methods on 1,449 programs in the test set and list the results in Table \ref{tab:res_stu}.

When reporting only 1 suspicious buggy line, \modelname successfully locates bugs in 421 (29.05\%) programs. The accuracy of the previous SOTA deep model, NBL, is only 20.29\% (9\% lower). \modelname also outperforms the three traditional methods significantly. For top-5 and top-10 accuracy, the superiority of \modelname is still clear. \modelname gives SOTA performance on student programs collected from real world.

\begin{figure*}[t]
    \centering
    \includegraphics[width=\columnwidth]{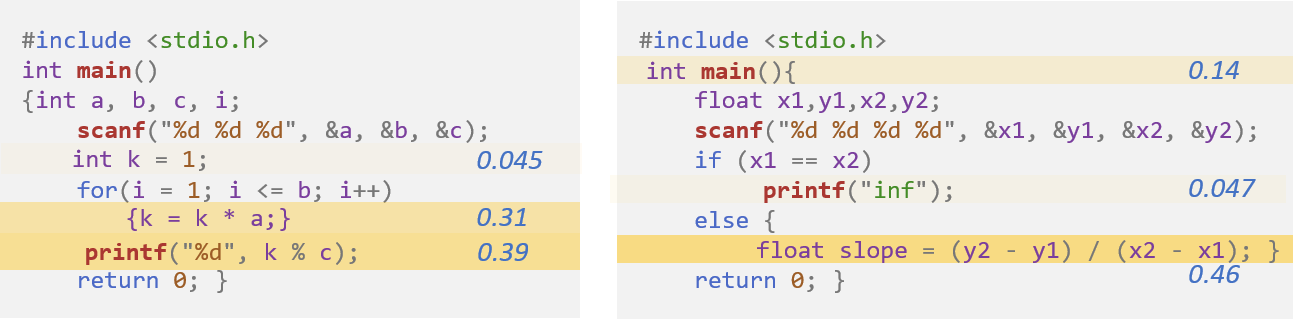}
    \caption{Visualization of \modelname on \Dstu. The three lines with highest importance scores are highlighted. The importance scores are labelled on the right with blue font.}
    \label{fig:well_stu}
\end{figure*}

\paragraph{Case study} Two programs in \Dstu are shown in Fig. \ref{fig:well_stu} with the importance scores given by \modelname. The first program is a solution for calculating $a^b\%c$, while the line ``k = k * a'' is buggy. The correct fix is ``k = k * a \% c''. \modelname ranks this line to the second place. The second figure is a program for calculating the slope of a line. However, the student forgets to write the ``printf'' statement to output the value of slope. \modelname locates the position to fix the bug (\ie, add the ``printf'' statement) at the hightlighted line successfully.

\subsection{Ablation on Attention Heads}
\label{sec:experiments:RQ3}

\begin{figure}[t]
    \centering
    \subfigure[\Dvar]{
        \includegraphics[width=0.42\columnwidth]{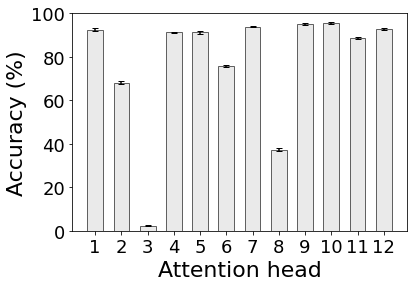}
        \label{fig:heads_var}
    }
    \subfigure[\Dbound]{
        \includegraphics[width=0.42\columnwidth]{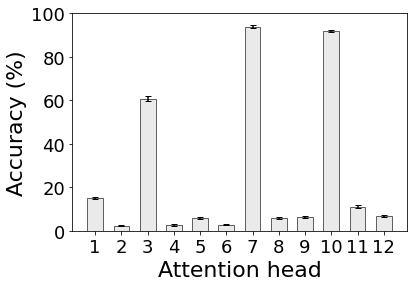}
        \label{fig:heads_bound}
    }
    \caption{Localization accuracy of \modelnameL{H$_i$} (attention head $i$). The histogram is the average of the 5 repeated trials, and the standard deviation is marked at the top of the histogram.} 
    \label{fig:heads}
\end{figure}

As aforementioned, we find that \modelname abnormally loses 23.32\% bug localization accuracy compared with \modelnameE in \Dbound. It may be caused by the issue of multi-head attention discovered in previous work \cite{head_pruning,head_pruning2}.
We delve into this issue in the following paragraphs and justify the necessity of extension in \modelnameE.
To investigate the impact of different heads, we carry out an ablation study by randomly sampling 2,000 correctly classified buggy examples from the test set to evaluate each \modelnameL{H$_i$} in \Dvar and \Dbound.

The localization accuracy of each \modelnameL{H$_i$} is shown in Fig. \ref{fig:heads}. In \Dbound, only three heads (3, 7, 10) are effective ($>$50\%) for localization, while the rest are almost invalid. The results agree with the previous work that only a small subset of heads does the heavy lifting \cite{head_pruning,head_pruning2}. On the contrary, in \Dvar, almost all heads are beneficial for localization, and only a few (3, 8) are invalid. One possible reason is that the data size of \Dvar is much larger and \modelname learns to focus on bugs better. This ablation study explains why \modelname performs slightly better than \modelnameE in \Dvar, but fails in \Dbound. Because in \Dbound, only several heads are effective while the others cause counteraction to the average aggregation, leading to an accuracy drop in \modelname.

\paragraph{Case study} Fig. \ref{fig:head} presents a case visualization of \modelnameL{H$_i$} in \Dbound. The gray-scale of the background refers to the importance score of the corresponding token, and the red box indicates the buggy location. \modelnameL{H$_7$} is accurate in the demonstrated case, due to the validity of the 7-th head according to Fig. \ref{fig:heads_bound}. As for the invalid heads (2 and 11), the visualized cases are distracted and erroneously predicted. In addition, \modelnameL{H$_7$} is actually \modelnameE as the 7-th head is the most important among all 12 heads. Therefore, the case study further demonstrates the feasibility of the extension in \modelnameE.
Please refer to Appendix \ref{sec:appendix:head} for more visualized cases.

In conclusion, the ``few-specialized'' problem of multi-head attention is also identified in \modelname. The ablation study verifies the extension of \modelnameE, which leverages fine-grained annotations as validation to select the most important and specialized head among all heads.

\begin{figure}[t] 
    \centering
  \subfigure[Visualization of importance scores from \modelnameL{H$_2$}, H$_7$ and H$_{11}$ in \Dbound. The gray-scale of the background indicates the importance score of the corresponding token, and \modelnameL{H$_i$} predicts the token with darkest background as the buggy location. The red box refers to the ground-truth buggy location.]{ 

    \begin{minipage}[t]{0.9\textwidth}
        
        $H_{2\textcolor{white}{1}}$~
            \input{appendix/head/isPrompt2}
            \\\hrule
        $H_{7\textcolor{white}{1}}$~
            \input{appendix/head/isPrompt7}
            \\\hrule
        $H_{11}$~
            \input{appendix/head/isPrompt11}
            \\
            \label{fig:head}
    \end{minipage}
    }

  \subfigure[Visualization of \modelname-{lstm} in \Dbound. The gray-scale of the background indicates the importance score of the corresponding token, and the red box refers to the ground-truth buggy location.]{
        \begin{minipage}[t]{0.9\textwidth}
        \input{appendix/lstm/hasKey}
        \\\hrule
        \input{appendix/lstm/hasStarted}
        \\\hrule
        \input{appendix/lstm/isGeq}
        \\
        \label{fig:lstm_vis}
        \end{minipage}
    }
  \caption{Visualization of \modelname.}
\end{figure}

\subsection{Portability \& Transferability}
\label{sec:experiments:RQ4}

In the previous experiments, \modelname shows great performance on bug localization. However, the effectiveness of \modelname may come from the CodeBERT backbone rather than weak supervision. Therefore, we perform another ablation study, by applying weak supervision to the LSTM model, creating \modelnameL{lstm}. The backbone of \modelnameL{lstm} is a two-layer bi-directional LSTM. An attention layer \cite{wang2016lstmattn} is placed on the top of LSTM, and we use it to compute the importance score $v$ (no aggregation since it is single-headed). We train and evaluate \modelnameL{lstm} upon \Dbound in a weakly supervised manner.

\begin{figure}
\centering
\includegraphics[width=0.4\columnwidth]{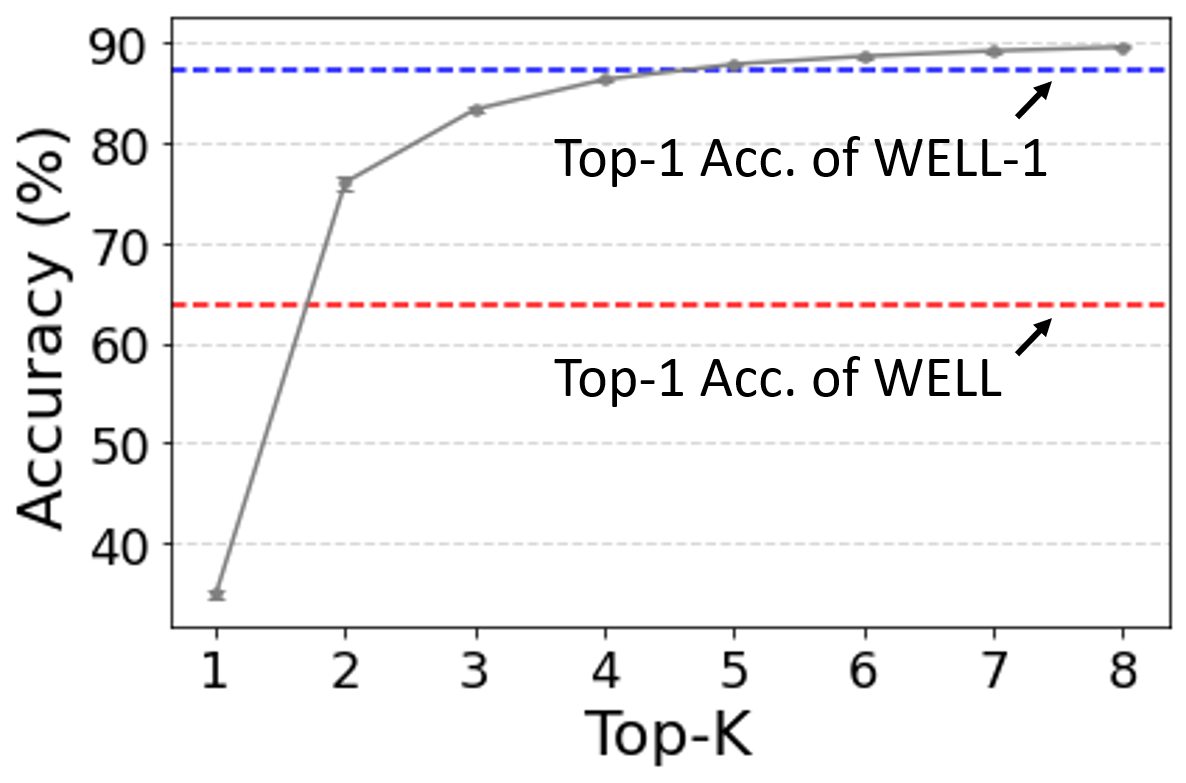}
\caption{Top-K localization accuracy curve with standard deviation of \modelnameL{lstm} in \Dbound. The standard deviation is too small and even imperceptible in the figure. The blue and red dashed lines are the top-1 localization accuracy of \modelnameE and \modelname respectively.}
\label{fig:topk}
\end{figure}

\paragraph{Top-K localization} The localization accuracy of \modelnameL{LSTM} is 34.96\%, which suggests that \modelnameL{lstm} is valid to locate \Dbound bugs although it is not as effective compared with CodeBERT and strongly supervised approaches.
The distracted attention may cause the rather not-effective-enough performance of WELL-lstm due to the insufficient model capability of LSTM. This is reasonable because the parameter size of LSTM (29M) is much smaller than CodeBERT (125M).
Therefore, to further demonstrate the potential of weak supervision, we also evaluate the top-K accuracy, where \modelnameL{lstm} selects the top-K important segments as the predicted buggy location, and any hit among them is considered correct. The results are shown in Fig. \ref{fig:topk}. The top-1 (exact) accuracy of \modelnameL{lstm} is 34.94\%, much lower than \modelname and \modelnameE, while the top-2 accuracy rises rapidly to 76.04\%. When we take four tokens into account, the accuracy is even 86.30\%, which is close to the exact accuracy of \modelnameE (87.19\%).

\paragraph{Case study} Some visualized cases are presented in Fig. \ref{fig:lstm_vis}, and more cases are shown in Appendix \ref{sec:appendix:lstm}.
Although the LSTM backbone is much less powerful than CodeBERT, \modelnameL{lstm} is still capable to notice the buggy locations in code in the scenario of weak supervision. As a rough conclusion, the weakly supervised framework is portable and transferable to other attention-based models such as LSTM as well.

Although the LSTM backbone is much less powerful than CodeBERT, \modelnameL{lstm} is still able to notice the buggy locations with only weak supervision. As a rough conclusion, the weakly supervised framework is portable and transferable to other attention-based models such as LSTM as well.

\subsection{Threats to Validity}
\label{sec:experiments:threat}

The randomness is a potential threat to the results. We counteract it by repeating the experiments in \textbf{RQ3} and \textbf{RQ4}.
The model selection could be a threat. We counteract it by comparing \modelname with the currently SOTA GREAT and CuBERT models. And we further build \modelnameL{$S$} with the controlled variable of supervision to compare with \modelname.
Also, the task and dataset selection may be another threat to validity. We counteract it by using two previously adopted benchmarks and one dataset constructed in this work. The subject datasets contain various of bugs and multiple programming languages.
The CodeBERT backbone may be too powerful and the weak supervision could be ineffective for other models, which makes a major threat to this paper too. We counteract it by replacing the CodeBERT backbone with LSTM in \textbf{RQ4} to demonstrate the impact of weak supervision.

\section{Conclusion \& Discussion}
\label{sec:conclusion}

This paper proposes \modelname, a weakly supervised bug localization model equipped with the powerful CodeBERT model, to alleviate the challenge of data collection and annotation.
\modelname is obtained on bug detection datasets without buggy-location annotations, making full usage of the more easily accessible training data.
Through the in-depth evaluations, we demonstrate that \modelname is capable of localizing bugs under coarse-grained supervision, and produces competitive or even better performance than existing SOTA models across both synthetic and real-world datasets. Further experiments show that the weakly supervised methodology in \modelname can be effectively applied to other attention-based models.

As an early step in this field, we hope this work could introduce some new ideas and methodologies into the SE community. Some future work is discussed in Appendix \ref{sec:appendix:fw}.

\bibliography{wileyNJD-AMA}

\begin{thebibliography}{10}
\providecommand \doibase [0]{http://dx.doi.org/}%

\bibitem{Mou2016Convolutional}
Mou L, Li G, Zhang L, Wang T, Jin Z. Convolutional Neural Networks over Tree
  Structures for Programming Language Processing. In:  Schuurmans D, Wellman
  MP. \kern-2pt, eds. {\it Proceedings of the Thirtieth {AAAI} Conference on
  Artificial Intelligence, February 12-17, 2016, Phoenix, Arizona, {USA}}{AAAI}
  Press; 2016\string: 1287--1293.

\bibitem{zhang2019novel}
Zhang J, Wang X, Zhang H, Sun H, Wang K, Liu X. A novel neural source code
  representation based on abstract syntax tree. In:  Atlee JM, Bultan T,
  Whittle J. \kern-2pt, eds. {\it Proceedings of the 41st International
  Conference on Software Engineering, {ICSE} 2019, Montreal, QC, Canada, May
  25-31, 2019}{IEEE} / {ACM}; 2019\string: 783--794.

\bibitem{yu2019neural}
Yu H, Lam W, Chen L, Li G, Xie T, Wang Q. Neural detection of semantic code
  clones via tree-based convolution. In:  Gu{\'{e}}h{\'{e}}neuc Y, Khomh F,
  Sarro F. \kern-2pt, eds. {\it Proceedings of the 27th International
  Conference on Program Comprehension, {ICPC} 2019, Montreal, QC, Canada, May
  25-31, 2019}{IEEE} / {ACM}; 2019\string: 70--80.

\bibitem{wang2020clone}
Wang W, Li G, Ma B, Xia X, Jin Z. Detecting Code Clones with Graph Neural
  Network and Flow-Augmented Abstract Syntax Tree. In:  Kontogiannis K, Khomh
  F, Chatzigeorgiou A, Fokaefs M, Zhou M. \kern-2pt, eds. {\it 27th {IEEE}
  International Conference on Software Analysis, Evolution and Reengineering,
  {SANER} 2020, London, ON, Canada, February 18-21, 2020}{IEEE}; 2020\string:
  261--271.

\bibitem{allamanis2016convolutional}
Allamanis M, Peng H, Sutton CA. A Convolutional Attention Network for Extreme
  Summarization of Source Code. In:  Balcan M, Weinberger KQ. \kern-2pt, eds.
  {\it Proceedings of the 33nd International Conference on Machine Learning,
  {ICML} 2016, New York City, NY, USA, June 19-24, 2016}. 48 of {\it {JMLR}
  Workshop and Conference Proceedings}. JMLR.org; 2016\string: 2091--2100.

\bibitem{code2vec}
Alon U, Zilberstein M, Levy O, Yahav E. code2vec: learning distributed
  representations of code. {\it Proc. {ACM} Program. Lang.} 2019\string;
  3({POPL})\string: 40:1--40:29.

\bibitem{li2017code}
Li J, Wang Y, King I, Lyu MR. Code Completion with Neural Attention and Pointer
  Networks. {\it CoRR} 2017\string; abs/1711.09573.

\bibitem{liu2020self}
Liu F, Li G, Wei B, Xia X, Fu Z, Jin Z. A Self-Attentional Neural Architecture
  for Code Completion with Multi-Task Learning. In: {ACM}; 2020\string: 37--47.

\bibitem{liu2020multi}
Liu F, Li G, Zhao Y, Jin Z. Multi-task Learning based Pre-trained Language
  Model for Code Completion. In: {IEEE}; 2020\string: 473--485.

\bibitem{hu2018deep}
Hu X, Li G, Xia X, Lo D, Jin Z. Deep code comment generation. In:  Khomh F, Roy
  CK, Siegmund J. \kern-2pt, eds. {\it Proceedings of the 26th Conference on
  Program Comprehension, {ICPC} 2018, Gothenburg, Sweden, May 27-28,
  2018}{ACM}; 2018\string: 200--210.

\bibitem{hu2018summarizing}
Hu X, Li G, Xia X, Lo D, Lu S, Jin Z. Summarizing Source Code with Transferred
  {API} Knowledge. In:  Lang J. \kern-2pt, ed. {\it Proceedings of the
  Twenty-Seventh International Joint Conference on Artificial Intelligence,
  {IJCAI} 2018, July 13-19, 2018, Stockholm, Sweden}ijcai.org; 2018\string:
  2269--2275.

\bibitem{code2seq}
Alon U, Brody S, Levy O, Yahav E. code2seq: Generating Sequences from
  Structured Representations of Code. In: OpenReview.net; 2019.

\bibitem{allamanis2018learning}
Allamanis M, Brockschmidt M, Khademi M. Learning to Represent Programs with
  Graphs. In: OpenReview.net; 2018.

\bibitem{vasic2019neural}
Vasic M, Kanade A, Maniatis P, Bieber D, Singh R. Neural Program Repair by
  Jointly Learning to Localize and Repair. In: OpenReview.net; 2019.

\bibitem{vincent2020global}
Hellendoorn VJ, Sutton C, Singh R, Maniatis P, Bieber D. Global Relational
  Models of Source Code. In: OpenReview.net; 2020.

\bibitem{kanade2020learning}
Kanade A, Maniatis P, Balakrishnan G, Shi K. Learning and Evaluating Contextual
  Embedding of Source Code. In: . 119 of {\it Proceedings of Machine Learning
  Research}. {PMLR}; 2020\string: 5110--5121.

\bibitem{benton2019defexts}
Benton S, Ghanbari A, Zhang L. Defexts: a curated dataset of reproducible
  real-world bugs for modern {JVM} languages. In:  Atlee JM, Bultan T, Whittle
  J. \kern-2pt, eds. {\it Proceedings of the 41st International Conference on
  Software Engineering: Companion Proceedings, {ICSE} 2019, Montreal, QC,
  Canada, May 25-31, 2019}{IEEE} / {ACM}; 2019\string: 47--50.

\bibitem{lutellier2020coconut}
Lutellier T, Pham HV, Pang L, Li Y, Wei M, Tan L. CoCoNuT: combining
  context-aware neural translation models using ensemble for program repair.
  In:  Khurshid S, Pasareanu CS. \kern-2pt, eds. {\it {ISSTA} '20: 29th {ACM}
  {SIGSOFT} International Symposium on Software Testing and Analysis, Virtual
  Event, USA, July 18-22, 2020}{ACM}; 2020\string: 101--114.

\bibitem{feng2020codebert}
Feng Z, Guo D, Tang D, et al. CodeBERT: {A} Pre-Trained Model for Programming
  and Natural Languages. In:  Cohn T, He Y, Liu Y. \kern-2pt, eds. {\it
  Proceedings of the 2020 Conference on Empirical Methods in Natural Language
  Processing: Findings, {EMNLP} 2020, Online Event, 16-20 November
  2020}Association for Computational Linguistics; 2020\string: 1536--1547.

\bibitem{wang2016defect}
Wang S, Liu T, Tan L. Automatically learning semantic features for defect
  prediction. In:  Dillon LK, Visser W, Williams LA. \kern-2pt, eds. {\it
  Proceedings of the 38th International Conference on Software Engineering,
  {ICSE} 2016, Austin, TX, USA, May 14-22, 2016}{ACM}; 2016\string: 297--308.

\bibitem{choi2017overruns}
Choi M, Jeong S, Oh H, Choo J. End-to-End Prediction of Buffer Overruns from
  Raw Source Code via Neural Memory Networks. In:  Sierra C. \kern-2pt, ed.
  {\it Proceedings of the Twenty-Sixth International Joint Conference on
  Artificial Intelligence, {IJCAI} 2017, Melbourne, Australia, August 19-25,
  2017}ijcai.org; 2017\string: 1546--1553.

\bibitem{li2018vulner}
Li Z, Zou D, Xu S, et al. VulDeePecker: {A} Deep Learning-Based System for
  Vulnerability Detection. In: The Internet Society; 2018.

\bibitem{pradel2018deepbugs}
Pradel M, Sen K. DeepBugs: a learning approach to name-based bug detection.
  {\it Proc. {ACM} Program. Lang.} 2018\string; 2({OOPSLA})\string:
  147:1--147:25.

\bibitem{zzh2017weak}
Zhou ZH. {A brief introduction to weakly supervised learning}. {\it National
  Science Review} 2017\string; 5(1)\string: 44-53.

\bibitem{dai2015boxsup}
Dai J, He K, Sun J. BoxSup: Exploiting Bounding Boxes to Supervise
  Convolutional Networks for Semantic Segmentation. In: {IEEE} Computer
  Society; 2015\string: 1635--1643.

\bibitem{papandreou2015weakly}
Papandreou G, Chen L, Murphy K, Yuille AL. Weakly- and Semi-Supervised Learning
  of a {DCNN} for Semantic Image Segmentation. {\it CoRR} 2015\string;
  abs/1502.02734.

\bibitem{lin2016scribblesup}
Lin D, Dai J, Jia J, He K, Sun J. ScribbleSup: Scribble-Supervised
  Convolutional Networks for Semantic Segmentation. In: {IEEE} Computer
  Society; 2016\string: 3159--3167.

\bibitem{bearman2016whats}
Bearman AL, Russakovsky O, Ferrari V, Li F. What's the Point: Semantic
  Segmentation with Point Supervision. In:  Leibe B, Matas J, Sebe N, Welling
  M. \kern-2pt, eds. {\it Computer Vision - {ECCV} 2016 - 14th European
  Conference, Amsterdam, The Netherlands, October 11-14, 2016, Proceedings,
  Part {VII}}. 9911 of {\it Lecture Notes in Computer Science}. Springer;
  2016\string: 549--565.

\bibitem{gap}
Lin M, Chen Q, Yan S. Network In Network. In:  Bengio Y, LeCun Y. \kern-2pt,
  eds. {\it 2nd International Conference on Learning Representations, {ICLR}
  2014, Banff, AB, Canada, April 14-16, 2014, Conference Track Proceedings};
  2014.

\bibitem{cam}
Zhou B, Khosla A, Lapedriza {\`{A}}, Oliva A, Torralba A. Learning Deep
  Features for Discriminative Localization. In: {IEEE} Computer Society;
  2016\string: 2921--2929.

\bibitem{gradcam}
Selvaraju RR, Cogswell M, Das A, Vedantam R, Parikh D, Batra D. Grad-CAM:
  Visual Explanations from Deep Networks via Gradient-Based Localization. In:
  {IEEE} Computer Society; 2017\string: 618--626.

\bibitem{wei2018revisiting}
Wei Y, Xiao H, Shi H, Jie Z, Feng J, Huang TS. Revisiting Dilated Convolution:
  {A} Simple Approach for Weakly- and Semi-Supervised Semantic Segmentation.
  In: {IEEE} Computer Society; 2018\string: 7268--7277.

\bibitem{ni2017weakly}
Ni J, Dinu G, Florian R. Weakly Supervised Cross-Lingual Named Entity
  Recognition via Effective Annotation and Representation Projection. In:
  Barzilay R, Kan M. \kern-2pt, eds. {\it Proceedings of the 55th Annual
  Meeting of the Association for Computational Linguistics, {ACL} 2017,
  Vancouver, Canada, July 30 - August 4, Volume 1: Long Papers}Association for
  Computational Linguistics; 2017\string: 1470--1480.

\bibitem{patra2019weakly}
Patra B, Moniz JRA. Weakly Supervised Attention Networks for Entity
  Recognition. In:  Inui K, Jiang J, Ng V, Wan X. \kern-2pt, eds. {\it
  Proceedings of the 2019 Conference on Empirical Methods in Natural Language
  Processing and the 9th International Joint Conference on Natural Language
  Processing, {EMNLP-IJCNLP} 2019, Hong Kong, China, November 3-7,
  2019}Association for Computational Linguistics; 2019\string: 6267--6272.

\bibitem{lison2020named}
Lison P, Barnes J, Hubin A, Touileb S. Named Entity Recognition without
  Labelled Data: {A} Weak Supervision Approach. In:  Jurafsky D, Chai J,
  Schluter N, Tetreault JR. \kern-2pt, eds. {\it Proceedings of the 58th Annual
  Meeting of the Association for Computational Linguistics, {ACL} 2020, Online,
  July 5-10, 2020}Association for Computational Linguistics; 2020\string:
  1518--1533.

\bibitem{safranchik2020weakly}
Safranchik E, Luo S, Bach SH. Weakly Supervised Sequence Tagging from Noisy
  Rules. In: {AAAI} Press; 2020\string: 5570--5578.

\bibitem{lei2016rationale}
Lei T, Barzilay R, Jaakkola TS. Rationalizing Neural Predictions. In:  Su J,
  Carreras X, Duh K. \kern-2pt, eds. {\it Proceedings of the 2016 Conference on
  Empirical Methods in Natural Language Processing, {EMNLP} 2016, Austin,
  Texas, USA, November 1-4, 2016}The Association for Computational Linguistics;
  2016\string: 107--117.

\bibitem{bastings2019rationale}
Bastings J, Aziz W, Titov I. Interpretable Neural Predictions with
  Differentiable Binary Variables. In:  Korhonen A, Traum DR, M{\`{a}}rquez L.
  \kern-2pt, eds. {\it Proceedings of the 57th Conference of the Association
  for Computational Linguistics, {ACL} 2019, Florence, Italy, July 28- August
  2, 2019, Volume 1: Long Papers}Association for Computational Linguistics;
  2019\string: 2963--2977.

\bibitem{yu2019complement}
Yu M, Chang S, Zhang Y, Jaakkola TS. Rethinking Cooperative Rationalization:
  Introspective Extraction and Complement Control. In:  Inui K, Jiang J, Ng V,
  Wan X. \kern-2pt, eds. {\it Proceedings of the 2019 Conference on Empirical
  Methods in Natural Language Processing and the 9th International Joint
  Conference on Natural Language Processing, {EMNLP-IJCNLP} 2019, Hong Kong,
  China, November 3-7, 2019}Association for Computational Linguistics;
  2019\string: 4092--4101.

\bibitem{deyoung2020eraser}
DeYoung J, Jain S, Rajani NF, et al. {ERASER:} {A} Benchmark to Evaluate
  Rationalized {NLP} Models. In:  Jurafsky D, Chai J, Schluter N, Tetreault JR.
  \kern-2pt, eds. {\it Proceedings of the 58th Annual Meeting of the
  Association for Computational Linguistics, {ACL} 2020, Online, July 5-10,
  2020}Association for Computational Linguistics; 2020\string: 4443--4458.

\bibitem{Jain2020bert}
Jain S, Wiegreffe S, Pinter Y, Wallace BC. Learning to Faithfully Rationalize
  by Construction. In:  Jurafsky D, Chai J, Schluter N, Tetreault JR.
  \kern-2pt, eds. {\it Proceedings of the 58th Annual Meeting of the
  Association for Computational Linguistics, {ACL} 2020, Online, July 5-10,
  2020}Association for Computational Linguistics; 2020\string: 4459--4473.

\bibitem{bahdanau2015attention}
Bahdanau D, Cho K, Bengio Y. Neural Machine Translation by Jointly Learning to
  Align and Translate. In:  Bengio Y, LeCun Y. \kern-2pt, eds. {\it 3rd
  International Conference on Learning Representations, {ICLR} 2015, San Diego,
  CA, USA, May 7-9, 2015, Conference Track Proceedings}; 2015.

\bibitem{vaswani2017transformer}
Vaswani A, Shazeer N, Parmar N, et al. Attention is All you Need. In:  Guyon I,
  Luxburg vU, Bengio S, et al. \kern-2pt, eds. {\it Advances in Neural
  Information Processing Systems 30: Annual Conference on Neural Information
  Processing Systems 2017, December 4-9, 2017, Long Beach, CA, {USA}};
  2017\string: 5998--6008.

\bibitem{luong2015effective}
Luong T, Pham H, Manning CD. Effective Approaches to Attention-based Neural
  Machine Translation. In:  M{\`{a}}rquez L, Callison{-}Burch C, Su J, Pighin
  D, Marton Y. \kern-2pt, eds. {\it Proceedings of the 2015 Conference on
  Empirical Methods in Natural Language Processing, {EMNLP} 2015, Lisbon,
  Portugal, September 17-21, 2015}The Association for Computational
  Linguistics; 2015\string: 1412--1421.

\bibitem{resnet}
He K, Zhang X, Ren S, Sun J. Deep Residual Learning for Image Recognition. In:
  {IEEE} Computer Society; 2016\string: 770--778.

\bibitem{layer_norm}
Ba LJ, Kiros JR, Hinton GE. Layer Normalization. {\it CoRR} 2016\string;
  abs/1607.06450.

\bibitem{bert}
Devlin J, Chang M, Lee K, Toutanova K. {BERT:} Pre-training of Deep
  Bidirectional Transformers for Language Understanding. In:  Burstein J, Doran
  C, Solorio T. \kern-2pt, eds. {\it Proceedings of the 2019 Conference of the
  North American Chapter of the Association for Computational Linguistics:
  Human Language Technologies, {NAACL-HLT} 2019, Minneapolis, MN, USA, June
  2-7, 2019, Volume 1 (Long and Short Papers)}Association for Computational
  Linguistics; 2019\string: 4171--4186.

\bibitem{roberta}
Liu Y, Ott M, Goyal N, et al. RoBERTa: {A} Robustly Optimized {BERT}
  Pretraining Approach. {\it CoRR} 2019\string; abs/1907.11692.

\bibitem{codesearchnet}
Husain H, Wu H, Gazit T, Allamanis M, Brockschmidt M. CodeSearchNet Challenge:
  Evaluating the State of Semantic Code Search. {\it CoRR} 2019\string;
  abs/1909.09436.

\bibitem{codexglue}
Lu S, Guo D, Ren S, et al. CodeXGLUE: A Machine Learning Benchmark Dataset for
  Code Understanding and Generation. {\it arXiv preprint arXiv:2102.04664}
  2021.

\bibitem{bpe}
Sennrich R, Haddow B, Birch A. Neural Machine Translation of Rare Words with
  Subword Units. In: The Association for Computer Linguistics; 2016.

\bibitem{head_pruning}
Voita E, Talbot D, Moiseev F, Sennrich R, Titov I. Analyzing Multi-Head
  Self-Attention: Specialized Heads Do the Heavy Lifting, the Rest Can Be
  Pruned. In:  Korhonen A, Traum DR, M{\`{a}}rquez L. \kern-2pt, eds. {\it
  Proceedings of the 57th Conference of the Association for Computational
  Linguistics, {ACL} 2019, Florence, Italy, July 28- August 2, 2019, Volume 1:
  Long Papers}Association for Computational Linguistics; 2019\string:
  5797--5808.

\bibitem{head_pruning2}
Michel P, Levy O, Neubig G. Are Sixteen Heads Really Better than One?. In:
  Wallach HM, Larochelle H, Beygelzimer A, d'Alch{\'{e}}{-}Buc F, Fox EB,
  Garnett R. \kern-2pt, eds. {\it Advances in Neural Information Processing
  Systems 32: Annual Conference on Neural Information Processing Systems 2019,
  NeurIPS 2019, December 8-14, 2019, Vancouver, BC, Canada}; 2019\string:
  14014--14024.

\bibitem{gupta19attr}
Gupta R, Kanade A, Shevade SK. Neural Attribution for Semantic Bug-Localization
  in Student Programs. In: ; 2019\string: 11861--11871.

\bibitem{abreu06spectrum}
Abreu R, Zoeteweij P, Gemund vAJC. An Evaluation of Similarity Coefficients for
  Software Fault Localization. In: {IEEE} Computer Society; 2006\string:
  39--46.

\bibitem{jones03spectrum}
Jones JA, Harrold MJ, Stasko J. Visualization for Fault Localization. In: ;
  2003.

\bibitem{wang2016lstmattn}
Wang Z, Yang B. Attention-based Bidirectional Long Short-Term Memory Networks
  for Relation Classification Using Knowledge Distillation from {BERT}. In:
  {IEEE}; 2020\string: 562--568.

\end{thebibliography}

\appendix

\section{Generating Dataset \Dbound}
\label{sec:appendix:bounderror}

We randomly sample 20\% projects from the Github Java Corpus \footnote{https://groups.inf.ed.ac.uk/cup/javaGithub/}, and extract all the Java methods using TreeSitter \footnote{https://tree-sitter.github.io/tree-sitter}. The methods with more than 400 tokens are discarded as they are too long and complicated.
Then we use TreeSitter to parse the methods to get the Concrete Syntax Tree (CST), and locate the subtree with type ``binary expression'' to find binary expressions with operators ``$<$='', ``$>$='', $<$ and $>$.
The off-by-one bug is brought into the methods by replacing comparison operators (\eg, $<$$\Leftrightarrow$$<$=, ``$>$''$\Leftrightarrow$``$>$=''\footnote{The equal condition in the comparison is added or removed in the operator.}). \footnote{All the used dataset will be published.}

\section{Model Configuration}
We implement \modelname with Python3 based on the DL framework PyTorch (ver 1.7.1) and the transformers package (ver 3.4.0) \footnote{\url{https://huggingface.co/transformers/}}.

\modelname adopts the released base version of CodeBERT (CodeBERT-base) as the backbone\footnote{\url{https://huggingface.co/microsoft/codebert-base}}. The max length is fixed to 512.
We utilize the open-sourced GREAT model with the same configuration reported in the original paper and train it from scratch.
As for CuBERT, we reproduce the model by replacing the backbone model with CodeBERT. The reasons are as follows: \ding{182} The model size of CuBERT is (three times larger than CodeBERT). It is too large for our machine to finetune the model. \ding{183} Using the same CodeBERT backbone helps us to compare the method with \modelname better.
During finetuning, the learning rate is set to $4\times10^{-5}$, and $L_2$ regularization is adopted with the weight of 0.01.
In each experiment, the models are trained/finetuned for 6 epochs with the batch size of 64 and we select the checkpoint with the highest accuracy on the validation set. To emphasize the supervision, the buggy locations in the training set is accessible for GREAT, CuBERT, while blocked for \modelname (weak supervision).
For \modelnameE, we sample 1,000 fine-grained labeled examples from the validation dataset to measure the importance of each attention head, which is a very small amount (less than 1\% of the original training dataset size). For \modelnameL{LSTM}, we train a bi-directional LSTM model with the hidden size of 600 and time step of 400. The vocabulary size is 30,000 and the embedding width is 512. In the experiment of \Dstu, the results of baseline models (NeuralBugLocator, Turantula, Ochiai and Diff-based model) are reported as in the original paper of NBL.

\section{\modelname for Bug Fixing}

\modelname is trained on bug detection datasets and can be applied for both bug detection and localization. Furthermore, \modelname has the potential for unsupervised bug fixing. The backbone model of \modelname is CodeBERT, which is pre-trained with masked language model task. So CodeBERT can predict the probability of the masked original token. Thus, when \modelname predict a program to be buggy and find the bug location, we can mask the located buggy tokens and query CodeBERT to predict the original tokens. Theoretically, CodeBERT should recover the most probable and correct tokens. In this way, we may apply \modelname for bug fixing.

However, there are still many challenges to accomplish this rough idea. For example, it is hard to determine the number of tokens to query CodeBERT to generate, \ie, the number of ``mask'' to insert to the bug location.
As an early trial, we try this method on fixing bugs in \Dvar dataset, in which each bug corresponds to a misused variable. We randomly sample 1,000 functions and evaluate the repair accuracy. When we set the repair (sub)token number to 1, 65.9\% of bugs are fixed correctly. When up to 3 (sub)tokens are taken into consideration, the number of fixed bugs grows to 81.4\%.
Although \modelname can only repair simple bugs now, this introduces a new thought to achieve bug fixing.

\section{More Visualized Cases of \modelname}
\label{sec:appendix:well}

\begin{figure*}[ht]
    \centering
    \subfigure[add]{
        \includegraphics[width=0.25\columnwidth]{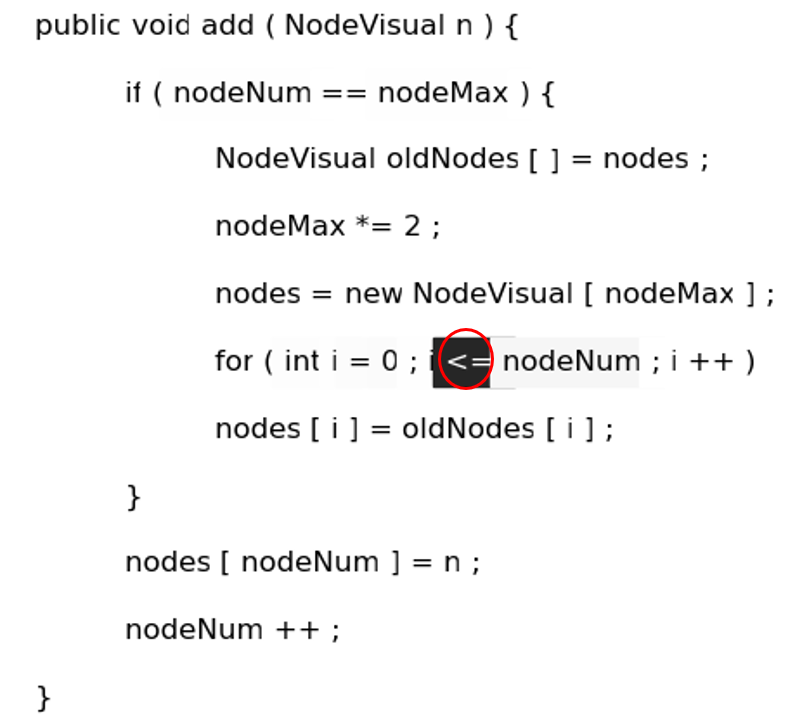}
        \label{fig:add}
    }
    \subfigure[clone]{
        \includegraphics[width=0.32\columnwidth]{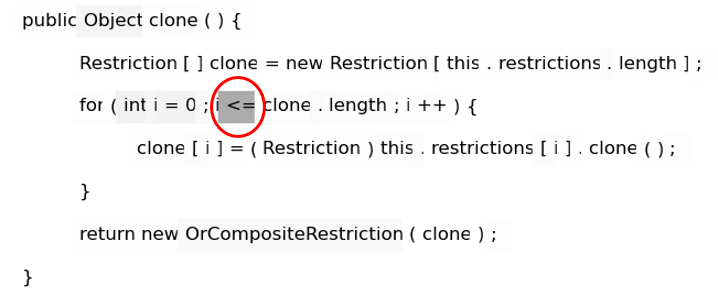}
        \label{fig:clone}
    }
    \subfigure[main]{
        \includegraphics[width=0.35\columnwidth]{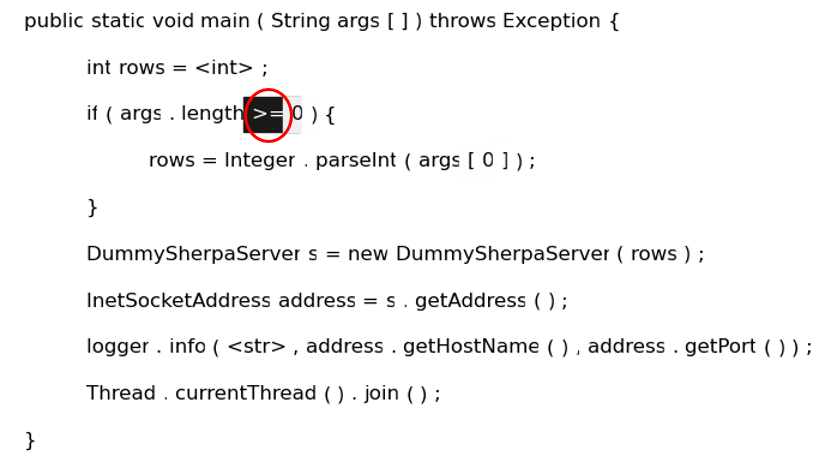}
        \label{fig:main}
    }
    \subfigure[delete]{
        \includegraphics[width=0.32\columnwidth]{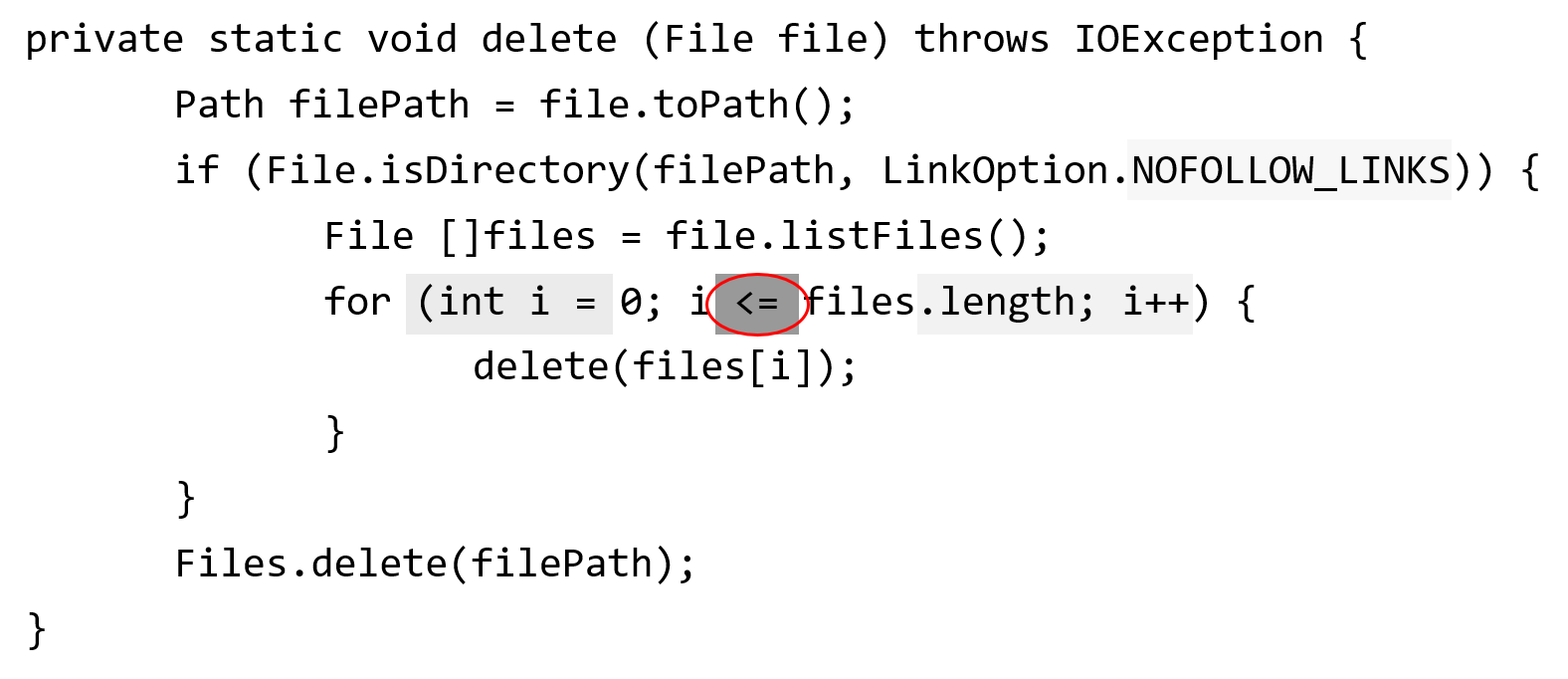}
        \label{fig:delete}
    }
    \subfigure[ReInitRounds]{
        \includegraphics[width=0.18\columnwidth]{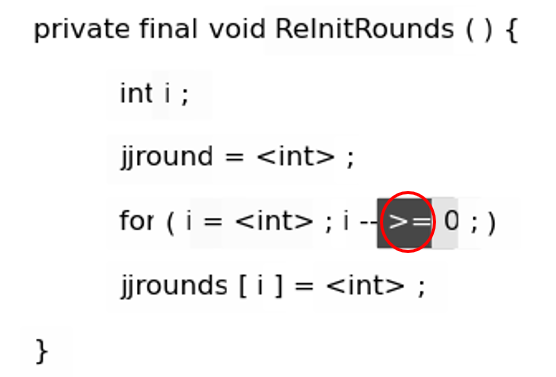}
        \label{fig:init}
    }
    \subfigure[setSample]{
        \includegraphics[width=0.42\columnwidth]{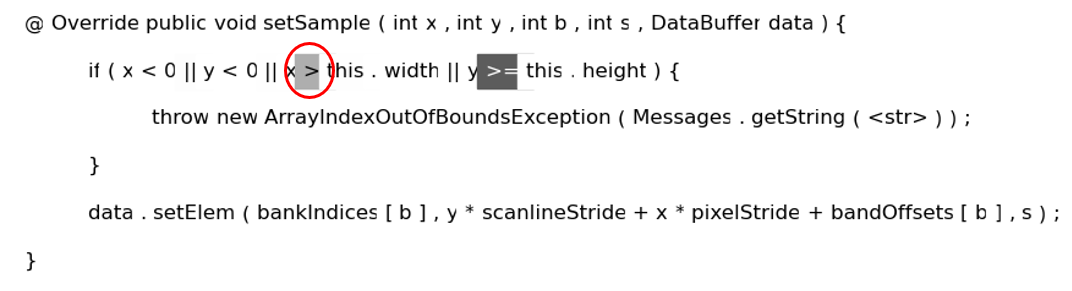}
        \label{fig:setsample}
    }
    \caption{Visualization of the importance score produced by \modelname in \Dbound. The red circle suggests the buggy location. The gray-scale of the background represents the importance score of the corresponding token. (a)-(e) are correctly handled by \modelname, as the darkest token is coincident with the ground-truth circle; while (f) is erroneously handled, as \modelname locate the bug at ``$>=$'' instead of ``$>$''.}
    \label{fig:well}
\end{figure*}

We list more visualized cases from \modelname in \Dbound in Fig. \ref{fig:well}. Perceivable dark backgrounds in each case are sparse, indicating that the importance scores are concentrated. Fig. \ref{fig:add}-\ref{fig:init} are correctly handled, and \modelname is quite ``certain'' about the buggy location predictions, as the importance scores are almost concentrated only upon the actual buggy locations. Although the bug in Fig. \ref{fig:setsample} is erroneously located by \modelname, it still pays attention to the actual buggy position, and is in a dilemma between ``$>$'' (actual bug) or ``$>=$'' (wrong prediction). In some certain scenarios, ``$>$'' may be misused; while in others, ``$>=$'' may be misused. As the context is not provided in this case, \modelname cannot make certain and correct decisions. We assume that it is hard to decide whether the equal condition should be incorporated given the limited context. And it is understandable for \modelname to make mistakes in such cases.

\begin{figure}[t]
    \centering
    \subfigure[hasFlag]{
            \begin{minipage}[*]{\columnwidth}
        $H_{2\textcolor{white}{1}}$~
            \input{appendix/head/hasFlag2}
            \\\hrule
        $H_{7\textcolor{white}{1}}$~
            \input{appendix/head/hasFlag7}
            \\\hrule
        $H_{11}$~
            \input{appendix/head/hasFlag11}
            \end{minipage}
            \label{fig:head:hasflag}
    }
    \subfigure[isDigit]{
            \begin{minipage}[*]{\columnwidth}
        $H_{2\textcolor{white}{1}}$~
            \input{appendix/head/isDigit2}
            \\\hrule
        $H_{7\textcolor{white}{1}}$~
            \input{appendix/head/isDigit7}
            \\\hrule
        $H_{11}$~
            \input{appendix/head/isDigit11}
            \end{minipage}
            \label{fig:head:isdigit}
    }
    \subfigure[previousPage]{
            \begin{minipage}[*]{\columnwidth}
        $H_{2\textcolor{white}{1}}$~
            \input{appendix/head/previousPage2}
            \\\hrule
        $H_{7\textcolor{white}{1}}$~
            \input{appendix/head/previousPage7}
            \\\hrule
        $H_{11}$~
            \input{appendix/head/previousPage11}
            \end{minipage}
            \label{fig:head:previouspage}
    }
    \subfigure[run]{
            \begin{minipage}[*]{\columnwidth}
        $H_{2\textcolor{white}{1}}$~
            \input{appendix/head/run2}
            \\\hrule
        $H_{7\textcolor{white}{1}}$~
            \input{appendix/head/run7}
            \\\hrule
        $H_{11}$~
            \input{appendix/head/run11}
            \end{minipage}
            \label{fig:head:run}
    }
    \subfigure[setFulfill]{
            \begin{minipage}[*]{\columnwidth}
        $H_{2\textcolor{white}{1}}$~
            \input{appendix/head/setFulfill2}
            \\\hrule
        $H_{7\textcolor{white}{1}}$~
            \input{appendix/head/setFulfill7}
            \\\hrule
        $H_{11}$~
            \input{appendix/head/setFulfill11}
            \end{minipage}
            \label{fig:head:setfulfill}
    }
    \subfigure[setMaxWriteThroughout]{
            \begin{minipage}[*]{\columnwidth}
        $H_{2\textcolor{white}{1}}$~
            \input{appendix/head/setMax2}
            \\\hrule
        $H_{7\textcolor{white}{1}}$~
            \input{appendix/head/setMax7}
            \\\hrule
        $H_{11}$~
            \input{appendix/head/setMax11}
            \end{minipage}
            \label{fig:head:setmax}
    }
    \caption{Visualization of the importance score produced by \modelnameL{H$_2$}, H$_7$ and H$_{11}$. The gray-scale of the background suggests the importance score of the corresponding token, and the red box refers to the buggy location. According to Fig. \ref{fig:heads_bound}, \modelnameL{H$_7$} is valid and effective and the other two are not.}
    \label{fig:head_more}
\end{figure}

\begin{figure}[t]
    \centering
    \begin{minipage}[*]{\columnwidth}
    \input{appendix/lstm/assertTag}
    \\\hrule
    \input{appendix/lstm/contained}
    \\\hrule
    \input{appendix/lstm/flagBit}
    \\\hrule
    \input{appendix/lstm/hasChildren}
    \\\hrule
    \input{appendix/lstm/isAcceptableSize}
    \end{minipage}
    \caption{Visualization of importance scores from \modelname-{lstm} in \Dbound. The gray-scale of the background suggests the importance score of the corresponding token, and the red box refers to the buggy location.}
    \label{fig:lstm:vis}
\end{figure}


\section{More Visualized Cases of \modelnameL{Hi}}
\label{sec:appendix:head}

We present more visualized cases of \modelnameL{H$_i$} on different attention heads in \Dbound in Fig. \ref{fig:head_more}. Similar to Fig. \ref{fig:head}, we carry out visualization upon three heads (\ie, 2, 7 and 11), where \modelnameL{H$_7$} is considered as effective and valid for bug localization while \modelnameL{H$_2$} and H$_{11}$ are not according to Fig. \ref{fig:heads_bound}. In all cases, \modelnameL{H$_7$} (\modelnameE) produces perceivable and concentrated importance score, and accurately locates the bugs. On the contrary, \modelnameL{H$_2$} and H$_{11}$ are distracted in many cases, and make less accurate localization. These cases to a certain extent demonstrates the issue of multi-head attention, and the necessity and feasibility of extented \modelnameE.

\section{More Visualized Cases of \modelnameL{lstm}}
\label{sec:appendix:lstm}

We provide more visualized cases of \modelnameL{lstm} in \Dbound in Fig. \ref{fig:lstm:vis}.
\modelnameL{lstm} exhibits similar behaviors in most cases, where the importance scores are concentrated on the buggy locations. However, there are also erroneous predictions, such as case 2. \modelnameL{lstm} regards ``segmentEnd'' instead of ``$<=$'' next to it as bug. Such incidents would somehow explain why \modelnameL{lstm} is not as effective as \modelname nor \modelnameE -- the LSTM backbone may be not strong enough to fully support the weakly supervised bug localization.


\section{Future Work}
\label{sec:appendix:fw}

As an early step in this field, we demonstrate the feasibility and effectiveness of weakly supervised bug localization, leaving many things to future work.

The evaluations of this paper are carried out on three synthesized datasets, where simple bugs are injected to the originally clean code to formulate buggy-location-annotated data. We leave the complex bug localization and the real world evaluation for future work.

The assumptions in \modelname, that the buggy fragment is consecutive and has a max length of $N$ may not be satisfied for many complex bugs. We could adopt the threshold activation strategy or the inconsecutive selection strategy to allow \modelname to locate inconsecutive token sequences with variable lengths. We leave this for the future work.

There are other approaches rather than attention to obtain the importance scores, such as gradients. \modelname may be generalized by employing such techniques. We leave these trials of different approaches to compute the importance scores for the future work.

A more desirable future work is to further fix bugs in the weakly supervised style. This could also be achieved with the CodeBERT backbone, when we consider not only $h_0$, but also other $h_i$'s. We leave this weakly supervised Dl-based bug fixing for the future work as well.

\end{document}